\UseRawInputEncoding
\documentclass[reprint,prl,aps]{revtex4-1}
\usepackage{graphicx}
\usepackage{color}
\usepackage{dcolumn}
\usepackage{bm}
\usepackage{amssymb}
\usepackage{color,amsmath}
\usepackage{soul,xcolor} 
\setstcolor{red} 
\usepackage{epstopdf}
\usepackage{hhline}
\usepackage{tabularx}

\newcolumntype{C}[1]{>{\centering\arraybackslash}m{#1}}
\newcolumntype{N}{@{}m{0pt}@{}}

\definecolor{cadmiumgreen}{rgb}{0.0, 0.42, 0.24}

\begin{document}

\title{Gate-tunable proximity effects in graphene on layered magnetic insulators}

\author{Chun-Chih Tseng$^{*1}$}
\author{Tiancheng Song$^{*1}$}
\author{Qianni Jiang$^{1}$}
\author{Zhong Lin$^{1}$}
\author{Chong Wang$^{2}$}
\author{Jaehyun Suh$^{2}$}
\author{K. Watanabe$^{3}$}
\author{T. Taniguchi$^{4}$}
\author{Michael A. McGuire$^{5}$}
\author{Di Xiao$^{1,2,6}$}
\author{Jiun-Haw Chu$^{1}$}
\author{David H. Cobden$^{1}$}
\author{Xiaodong Xu$^{1,2}$}
\author{Matthew Yankowitz$^{1,2\dagger}$}

\affiliation{$^{1}$Department of Physics, University of Washington, Seattle, Washington, 98195, USA}
\affiliation{$^{2}$Department of Materials Science and Engineering, University of Washington, Seattle, Washington, 98195, USA}
\affiliation{$^{3}$Research Center for Functional Materials, National Institute for Materials Science, 1-1 Namiki, Tsukuba 305-0044, Japan}
\affiliation{$^{4}$International Center for Materials Nanoarchitectonics, National Institute for Materials Science,  1-1 Namiki, Tsukuba 305-0044, Japan}
\affiliation{$^{5}$Materials Science and Technology Division, Oak Ridge National Laboratory, Oak Ridge, Tennessee, 37831, USA}
\affiliation{$^{6}$Pacific Northwest National Laboratory, Richland, WA, USA}
\affiliation{$^{*}$These authors contributed equally to this work.}
\affiliation{$^{\dagger}$myank@uw.edu (M.Y.)}

\maketitle

\textbf{The extreme versatility of two-dimensional van der Waals (vdW) materials derives from their ability to exhibit new electronic properties when assembled in proximity with dissimilar crystals~\cite{Geim2013}. For example, although graphene is inherently non-magnetic, recent work has reported a magnetic proximity effect in graphene interfaced with magnetic substrates~\cite{Wang2015PRLgrMPE,Wei2016,Leutenantsmeyer2016,Xu2018Natcom,Tang2018APLMat,Karpiak2019,Tang2020Adv,Wu2020NatEle,Ghiasi2021,Wu2021magnet,Chau2022npj}, potentially enabling a pathway towards achieving a high-temperature quantum anomalous Hall effect~\cite{Qiao2010PRB,Qiao2014PRL,Zhang2015PRB,Zhang2018PRB,Hog2020PRB}. Here, we investigate heterostructures of graphene and chromium trihalide magnetic insulators (CrI$_3$, CrBr$_3$, and CrCl$_3$). Surprisingly, we are unable to detect a magnetic exchange field in the graphene, but instead discover proximity effects featuring unprecedented gate-tunability. The graphene becomes highly hole-doped due to charge transfer from the neighboring magnetic insulator, and further exhibits a variety of atypical transport features. These include highly extended quantum Hall plateaus, abrupt reversals in the Landau level filling sequence, and hysteresis over at least days-long time scales. In the case of CrI$_3$, we are able to completely suppress the charge transfer and all attendant atypical transport effects by gating. The charge transfer can additionally be altered in a first-order phase transition upon switching the magnetic states of the nearest CrI$_3$ layers. Our results provide a roadmap for exploiting the magnetic proximity effect in graphene, and motivate further experiments with other magnetic insulators.}

Assembling heterostructures of van der Waals (vdW) crystals enables the creation of new properties that do not exist in the constituent materials alone. For example, combining proximity-induced magnetism~\cite{Wang2015PRLgrMPE,Wei2016,Leutenantsmeyer2016,Xu2018Natcom,Tang2018APLMat,Karpiak2019,Tang2020Adv,Wu2020NatEle,Ghiasi2021,Wu2021magnet,Chau2022npj} and spin-orbit coupling~\cite{Avsar2014NatCom,Wang2015NatCom,Wang2016PRX,Yang2016SOC,Island2019} in graphene has been a longstanding goal, as a high-temperature quantum anomalous Hall effect is predicted to arise in such a system~\cite{Qiao2010PRB,Qiao2014PRL,Zhang2015PRB,Zhang2018PRB,Hog2020PRB}. Chromium trihalides are a prototypical family of two-dimensional magnetic insulators~\cite{Huang2017CrI3,Zhang2019NanoLet,Cai2019NanoLet}, and are promising for realizing a variety of proximity effects when interfaced with graphene owing to their intrinsic ferromagnetic ordering, spin-orbit coupling, and large electron affinities. However, so far pristine interfaces between graphene and chromium trihalides have not been reported owing challenges arising from the extreme environmental sensitivity of the latter crystals.

\begin{figure*}[t]
\includegraphics[width=6.9 in]{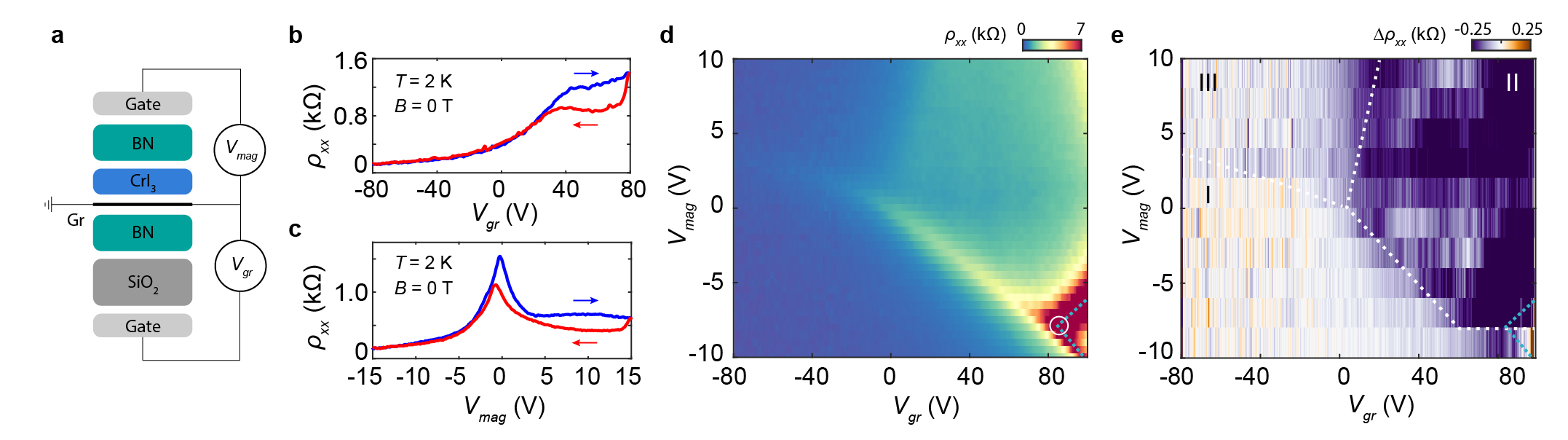} 
\caption{\textbf{Transport in graphene on CrI$_3$ at zero magnetic field.}
\textbf{a}, Schematic of the device structure. Monolayer graphene is interfaced with CrI$_3$ and encapsulated by BN. The voltage on the gate facing the graphene is $V_{gr}$, and the voltage on the gate facing the CrI$_3$ is $V_{mag}$. 
\textbf{b}, Four-terminal resistivity of a device with trilayer CrI$_3$ (Device A) as $V_{gr}$ is swept back and forth, with $V_{mag}=0$.
\textbf{c}, Resistivity as $V_{mag}$ is swept back and forth with $V_{gr}=0$.
\textbf{d}, Map of the device resistivity acquired by sweeping both gates. The trajectory of the Dirac point is denoted by the blue dashed curve, as determined by Hall effect measurements at $B=4$~T (see Supplementary Information Fig.~\ref{fig:DeviceA_DP4T}). The white circle denotes the point at which the trajectory of the Dirac point reverses.
\textbf{e}, Map of the transport hysteresis, $\Delta \rho_{xx}$, acquired by taking the difference between the resistivity upon slowly sweeping $V_{gr}$ forward and backward. The white dashed curve is a guide to the eye, corresponding approximately to resistive peaks and plateaus seen in \textbf{d}.
}
\label{fig:1}
\end{figure*}

Here, we report low-temperature transport measurements of graphene on thin substrates of CrI$_3$, CrBr$_3$, and CrCl$_3$ (collectively referred to as CrX$_3$). Figure~\ref{fig:1}a shows a schematic of the general device structure we fabricate. Interfaces of graphene and thin CrX$_3$ are encapsulated with boron nitride (BN) and surrounded by top and bottom gates. The CrX$_3$ crystals we use range from three to tens of layers in thickness, however, the majority of our results do not appear to depend meaningfully on this parameter. In order to avoid degradation of the CrX$_3$ crystals during device fabrication, we first shape a flake of exfoliated graphene into a Hall bar geometry using a polymer-free anodic oxidation technique with an atomic force microscope tip~\cite{Li2018AON}, and then assemble the entire vdW heterostructure in an argon-filled glovebox (see Methods and Supplementary Information Figs.~\ref{fig:device_images} and~\ref{fig:device_fab} for full details). We fabricate devices in which the graphene rests atop CrX$_3$, and visa versa, and see the same behavior in both cases. For clarity, we henceforth refer to the bias on the gate facing the graphene (CrX$_3$) as $V_{gr}$ ($V_{mag}$). 

We focus our attention primarily on graphene/CrI$_3$ heterostructures, from which we can additionally understand the salient properties of graphene/CrBr$_3$ and CrCl$_3$ (see Methods and Supplementary Figs.~\ref{fig:CrBr3_device}-\ref{fig:CrCl3_thin}). CrI$_3$ has the lowest electron affinity of the three chromium trihalides, and as a result the modulation doping of the graphene is the smallest. Figure~\ref{fig:1}b (c) shows the resistivity of a graphene on trilayer CrI$_3$ device (Device A) measured as $V_{gr}$ ($V_{mag}$) is swept back and forth with the other gate grounded at a temperature of $T=2$~K. We see a number of features that are uncharacteristic of pristine graphene encapsulated only with BN. First, the transport differs notably depending on which of the two gates is swept. Second, the transport is hysteretic, with the hysteresis most pronounced at positive values of either gate voltage. Although there are kinks or peaks in the resistivity suggestive of a Dirac point, measurements of the corresponding Hall resistance reveal that the graphene is hole-doped over most of the accessible gate voltage range (see Supplementary Information Fig.~\ref{fig:DeviceA_DP4T}), indicating that these resistive peaks arise from a different mechanism. 

Figure~\ref{fig:1}d shows a map of the device resistivity acquired by sweeping both of the gates. The blue dashed curve traces the position of the Dirac point as determined by Hall effect measurements (Supplementary Information Fig.~\ref{fig:DeviceA_DP4T}). The Dirac point evolves with the two gate voltages as anticipated from simple electrostatics in the bottom rightmost portion of the map. However, its trajectory abruptly reverses as the bias on $V_{mag}$ is further reduced towards zero (white circle in Fig.~\ref{fig:1}d). The bent trajectory of the Dirac point indicates a nonlinear and nonmonotonic relationship between the gate voltage and graphene charge carrier density, in stark contrast with the behavior of conventional monolayer graphene devices in which the gate capacitance is fixed. We see other transport features atypical of graphene, including an anomalous resistive peak that moves roughly diagonally across the map, as well as an abrupt resistive step separating the top left and right halves of the map. There is also a sharp resistivity increase in the top rightmost corner of the map that indicates the reappearance of the Dirac point. Further, these features are directly associated with the hysteretic graphene transport. Figure~\ref{fig:1}e shows a measurement of $\Delta \rho_{xx}$ acquired by taking the difference between $\rho_{xx}$ measured as $V_{gr}$ is slowly swept back and forth (see Supplementary Information Fig.~\ref{fig:DeviceA_reverse} for the reverse measurements). As a guide to the eye, the white dashed curve denotes the positions of the anomalous resistive peaks we observe in Fig.~\ref{fig:1}d, and separates the map into Regions I, II, and III. The hysteresis is most prominent in Region II of the map, approximately bounded by the anomalous resistive peaks and plateaus.

\begin{figure*}[t]
\includegraphics[width=5.5 in]{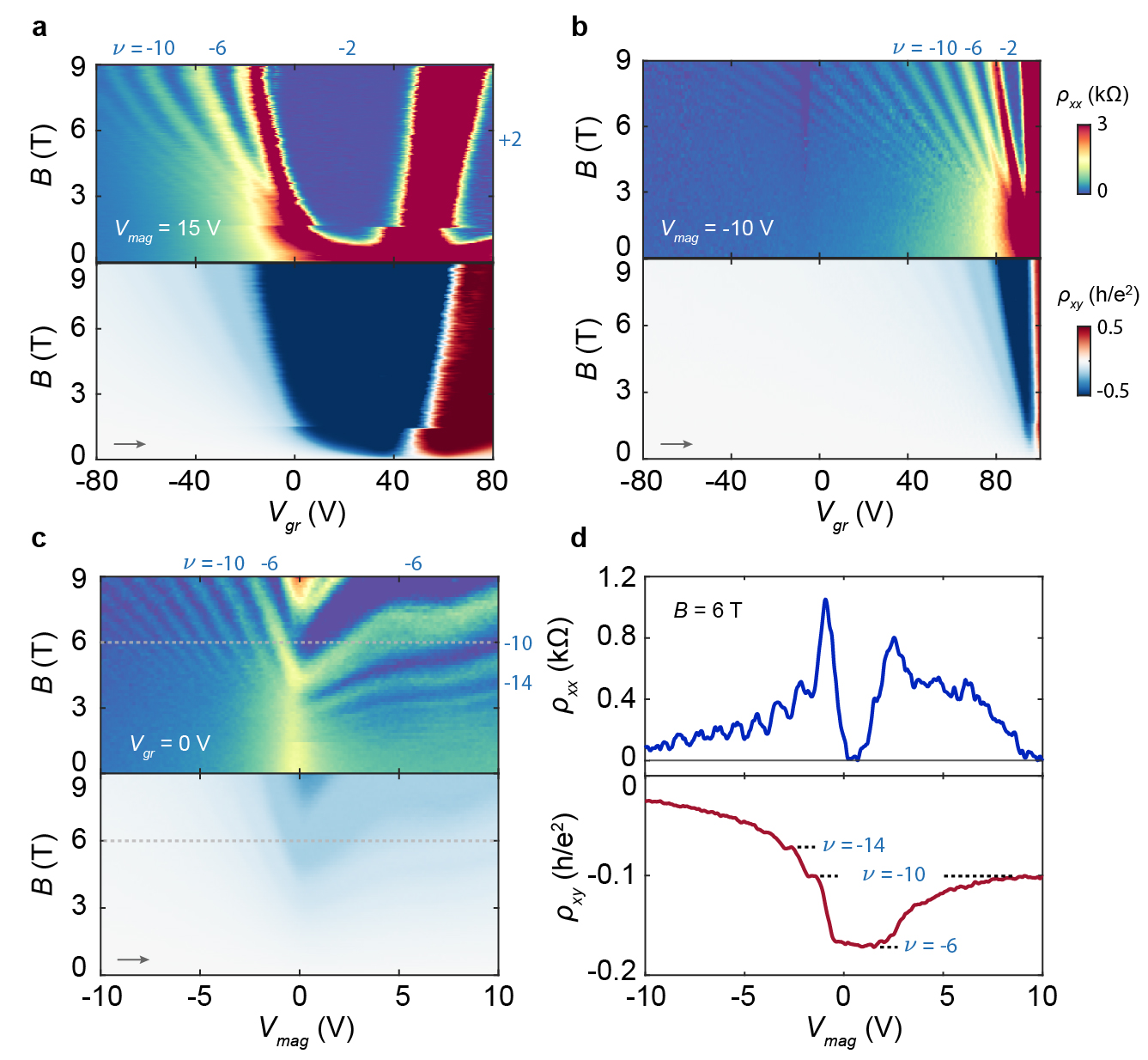} 
\caption{\textbf{Landau fan diagrams of graphene on CrI$_3$.}
\textbf{a-b}, (top) Longitudinal, $\rho_{xx}$, and (bottom) Hall, $\rho_{xy}$, resistivities of Device A acquired by sweeping $V_{gr}$ from negative to positive values with $V_{mag}=15$~V (\textbf{a}) and $V_{mag}=-10$~V (\textbf{b}), respectively.
\textbf{c}, Landau fan diagram acquired by sweeping $V_{mag}$ from negative to positive with $V_{gr}=0$.
\textbf{d}, $\rho_{xx}$ (top) and $\rho_{xy}$ (bottom) cuts from \textbf{c} acquired at $B=6$~T. 
}
\label{fig:2}
\end{figure*}

Transport measurements in a magnetic field, $B$, help to reveal the origin of these unusual features. Figures~\ref{fig:2}a-b show Landau fan diagrams of the longitudinal, $\rho_{xx}$ (top), and Hall, $\rho_{xy}$ (bottom), resistivities acquired by sweeping $V_{gr}$ from negative to positive bias with fixed values of $V_{mag}=15$~V and $-10$~V, respectively. The latter is consistent with typical hole-doped graphene: the Dirac point appears at large positive $V_{gr}$ and is associated with a sign change in $\rho_{xy}$ upon doping, and there is a series of integer quantum Hall (IQH) states that disperse linearly away from the Dirac point. These correspond to filling factors of $\nu=-2,-6,-10,...$, consistent with the usual sequence of states arising from spin- and valley-degenerate monolayer graphene Landau levels. In contrast, at $V_{mag}=15$~V we see a number of anomalous features in the Landau fan, including the Dirac point drifting with magnetic field, an abrupt resistivity jump at $B=1.6$~T, extremely wide $\nu=\pm2$ IQH plateaus, and IQH states at higher filling factors with slightly widened plateaus that move nonlinearly. 

Landau fans acquired by sweeping $V_{mag}$ at fixed $V_{gr}$ exhibit even more striking peculiarities. Figure~\ref{fig:2}c shows a representative example, in which the IQH states disperse as expected for $V_{mag} \lesssim 0$, but abruptly reverse direction for $V_{mag} \gtrsim 0$. The latter regime corresponds to an apparent negative compressibility of the system, in which applying more positive gate voltage results in filling additional hole-type Landau levels, rather than their anticipated depletion. Figure~\ref{fig:2}d shows a representative example of this phenomenon at $B=6$~T, in which the graphene exhibits two disconnected regimes of doping corresponding to the $\nu=-10$ IQH state. 

The IQH states in typical graphene devices fan out linearly from the Dirac point as the magnetic field is raised. Their trajectories are described by the St\v{r}eda formula~\cite{Streda1982}, $\nu=(h/e)(\partial n/\partial B)$, where $h$ is Planck's constant, $e$ is the charge of the electron, and $n$ is the charge carrier density. Departure from this behavior provides further evidence of the nonlinear relationship between the gate voltage and the charge carrier density in the graphene originating from the charge transfer with the CrI$_3$, consistent with the bent trajectory of the Dirac point observed in Fig.~\ref{fig:1}d. The wide IQH plateaus and their nonlinear trajectories in the fan diagram indicate that the charges induced by the gate do not accumulate in the graphene but rather fill the CrI$_3$, since electrons become localized in the insulating CrI$_3$ and do not contribute to transport. Related effects have been previously observed in graphene on SiC~\cite{Lafont2015,Alexander-Webber2016} and CrOCl~\cite{Wang2015gr-crocl}. The coexistence of normal and atypical quantum Hall effects within our device suggests that the modulation doping can be controlled by gating, and under suitable conditions is even suppressed entirely.

\begin{figure*}[t]
\includegraphics[width=6.9 in]{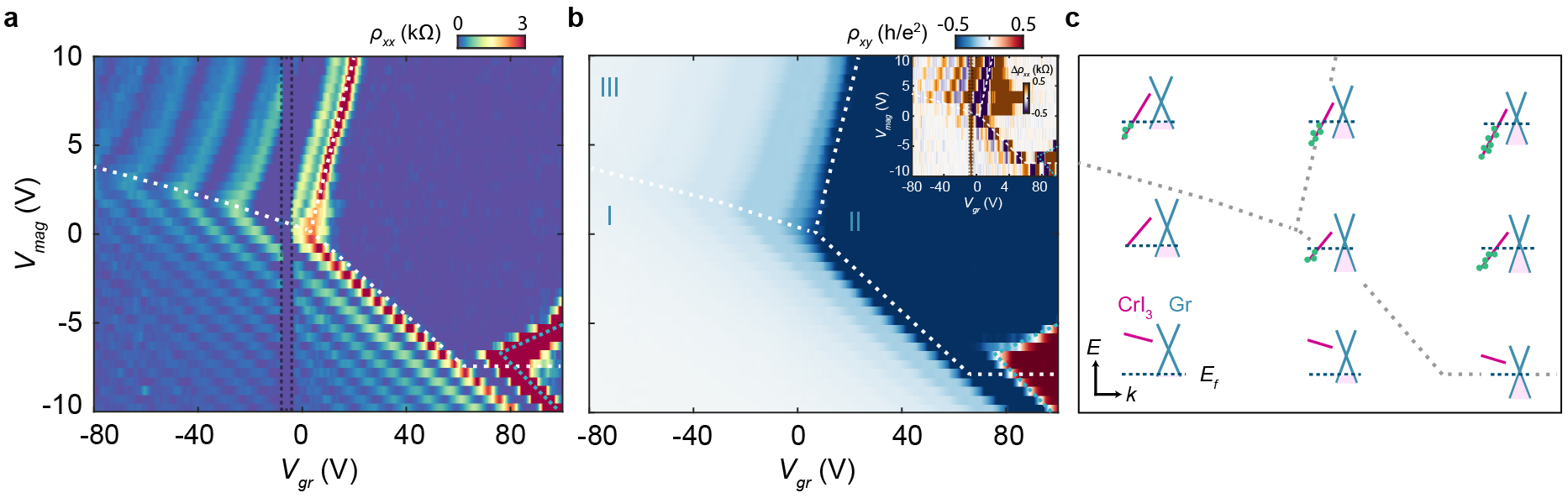} 
\caption{\textbf{Mapping the charge transfer and band alignment in graphene on CrI$_3$.}
\textbf{a-b}, Maps of $\rho_{xx}$ (\textbf{a}) and $\rho_{xy}$ (\textbf{b}) acquired by sweeping both gates at $B=9$~T. The white dotted lines are the same guides to the eye as in Fig.~\ref{fig:1}e, and separate the map into Regions I, II, and III. The $\nu=0$ state, corresponding to the Dirac point, is determined by the sign change in $\rho_{xy}$ (blue dashed curve). (inset of \textbf{b}) Map of the transport hysteresis, $\Delta \rho_{xx}$, acquired at $B=9$~T. The regions enclosed within the vertical black and gray boxes in \textbf{a} and the inset of \textbf{b} are contaminated by artifacts owing to insulating behavior at the contacts.
\textbf{c}, Inferred alignment of the graphene Dirac cone (blue) and the lowest electron-holding states in CrI$_3$ (red) under different gating conditions, corresponding to the approximate corresponding positions from the maps in \textbf{a-b}. Because the CrI$_3$ is multiple layers thick, gate voltages establish an electric field across the sheet which tilts the bands. The band offset at the interface between the graphene and adjacent CrI$_3$ layer is a fixed quantity determined by the graphene work function and CrI$_3$ electron affinity. The dark blue dashed line denotes the Fermi energy, $E_F$, in the graphene. Filled states in graphene are indicated in pink, and electrons in CrI$_3$ are denoted schematically by the green dots.
}
\label{fig:3}
\end{figure*}

We characterize the full dependence of the charge transfer on gating and magnetic field by measuring $\rho_{xx}$ and $\rho_{xy}$ over the entire accessible range of both gate voltages at a fixed $B=9$~T (Fig.~\ref{fig:3}a-b). For the purposes of analyzing the behavior, we divide the map into three regions as marked by the white dotted lines (reproduced from Fig.~\ref{fig:1}e). Region I exhibits typical graphene magnetotransport, in which a sequence of hole-doped IQH states disperses diagonally. The negative slope of these features is consistent with the ratio of the geometrical capacitances of the top and bottom gates. This remains true for the $\nu=0$ insulating state in the bottom right corner of the map, across which the sign of the Hall effect flips. In contrast, Region II corresponds almost entirely to the $\nu=-2$ IQH state. At the foot of Region II, we observe an abrupt reversal in the trajectory of the $\nu=0$ state similar to that of the Dirac point at $B=0$ (Fig.~\ref{fig:1}d). In Region III, the IQH states become nearly insensitive to $V_{mag}$ and develop a positive slope. The inset of Fig.~\ref{fig:3}b shows hysteresis measurements, $\Delta \rho_{xx}$, acquired at $B=9$~T, analogous to the zero-field map shown in Fig.~\ref{fig:1}e. The hysteresis is primarily confined to Region III; however, we note that this measurement scheme is largely insensitive to hysteresis in Region II owing to the extended $\nu=-2$ plateau. In combination with the zero-field measurements (Fig.~\ref{fig:1}e), we deduce that the device exhibits hysteresis in both Regions II and III. The hysteresis is therefore directly associated with regions of atypical graphene transport, and does not occur in Region I where the transport is conventional.

\begin{figure*}[t]
\includegraphics[width=6 in]{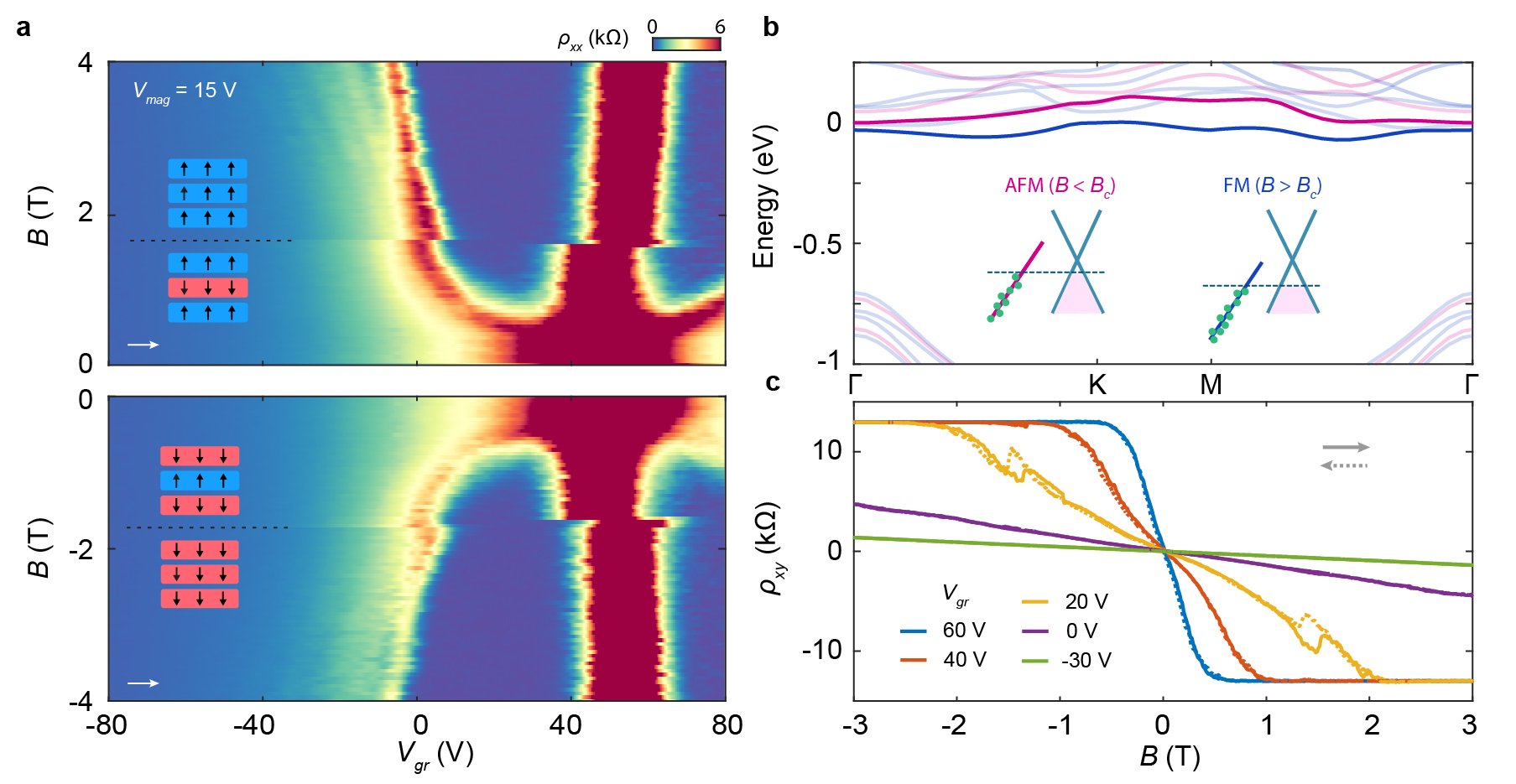} 
\caption{\textbf{Magnetic field-dependent modulation doping of graphene on CrI$_3$.}
\textbf{a}, Low-field Landau fan diagram acquired by sweeping $V_{gr}$ from negative to positive with $V_{mag}=15$~V. An abrupt shift in the modulation doping is denoted by the black dashed line, corresponding to the critical field at which the trilayer CrI$_3$ flips between interlayer AFM and FM (illustrated by the cartoon insets).
\textbf{b}, Ab initio calculation of the band structure of bilayer CrI$_3$ in the interlayer FM (blue) and AFM (red) states. The lowest conduction band is highlighted for clarity. The cartoon illustrates the additional modulation doping of graphene anticipated in the FM state compared with the AFM state owing to the lower conduction band energy.
\textbf{c}, $\rho_{xy}$ acquired by sweeping $B$ back and forth at different values of $V_{gr}$, with $V_{mag}=15$~V.
}
\label{fig:4}
\end{figure*}

The above measurements can all be qualitatively understood by taking into account the gate- and field-dependent charge density in the CrI$_3$. Figure~\ref{fig:3}c shows a series of cartoon diagrams that depict the electron states and their occupancies in the graphene (blue) and CrI$_3$ (red) for different combinations of the top and bottom gate voltages. Filled states in graphene below the chemical potential (blue dashed line) are colored pink. The red line represents the lowest-energy electron states in the CrI$_3$, and corresponds either to the bottom of the conduction band or a band of in-gap defect states; the latter is more likely because the charge mobility is very low. Electrons in the CrI$_3$ are indicated by the green dots. We assume that the alignment between the graphene Dirac point and the lowest-energy electron band in the neighboring CrI$_3$ layer is fixed by the combination of the graphene work function and CrI$_3$ electron affinity. Biasing $V_{mag}$ establishes an electric field across the few-layer CrI$_3$, shifting the energy of the CrI$_3$ states either up or down relative to the graphene. 

Within Region I, the chemical potential lies beneath the lowest-energy electron states throughout the CrI$_3$, and as a result they are all unoccupied. The CrI$_3$ then simply behaves as a dielectric, and changing the bias on either gate capacitively dopes the graphene as usual. In Regions II and III, the bias $V_{mag}$ is such that some electron states in the CrI$_3$ are below the chemical potential, causing electrons to tunnel into them from the graphene. These electrons become almost entirely localized and contribute negligibly to the conductivity, but are mobile enough in the out-of-plane direction to screen the underlying gate. This results in a greatly altered sensitivity of the graphene doping to changes in $V_{mag}$. The hysteresis observed in these regions results from an inability of the electrons to reach a true equilibrium due to long timescales in the CrI$_3$. As further evidence for this, we observe irreversible changes in the modulation-doping of the graphene in measurements performed days apart (see Supplementary Fig.~\ref{fig:long_drift}), and in a graphene/CrCl$_3$ device doped to a similar regime, the resistance exhibits telegraph noise on a timescale of tens of minutes when both gate voltages are held fixed (Supplementary Fig.~\ref{fig:telegraph}). 

The distinction between Regions II and III can be explained by the density of states in the graphene, which is small or vanishing in the former and much larger in the latter. These regions are separated by a resistivity plateau at zero field, and by a crossover at 9 T from the highly extended $\nu=-2$ IQH plateau to a sequence of less extended IQH states with higher filling factors. In Region II, changes in $V_{mag}$ are almost entirely screened. In this region, changing $V_{gr}$ only weakly dopes the graphene because the low density of states means that adding electrons to the graphene causes its electrochemical potential to rise rapidly, leading to more electrons tunneling into the CrI$_3$. In Region III, changing $V_{gr}$ dopes the graphene more strongly because its density of states is larger. Meanwhile, changing $V_{mag}$ does somewhat dope the graphene, but, strikingly, it does so in the wrong sense. In this case, the IQH states have positive slope indicative of an effectively negative differential capacitance. In other words, a more positive $V_{mag}$ results in larger hole doping of the graphene, whereas naively it would be expected to instead reduce the hole doping. This effect is also responsible for the peculiar reversal of the trajectories of the IQH plateaus seen near $V_{mag}=0$ in the Landau fans shown in Figs.~\ref{fig:2}c-d. Negative differential capacitance can result from negative compressibility in a strongly correlated conductor, such as has been reported for graphene on MoS$_2$~\cite{Larentis2014}. It can also result from  a large rearrangement of charge in the dielectric induced by a small change in applied electric field, such as occurs when a ferroelectric polarization flips. Rearrangement of the interacting electrons within the CrI$_3$ therefore seems the most likely explanation for this phenomenon, and is consistent with the associated hysteresis, though it appears too complex to be usefully modeled at this point.

So far, we have not considered the role of the magnetic ordering of the CrI$_3$, which exhibits out-of-plane intralayer ferromagnetism (FM) and antiferromagnet (AFM) interlayer ordering at low temperature. Figure~\ref{fig:4}a shows a low-field Landau fan diagram acquired for both positive and negative values of $B$. As noted earlier, upon increasing the field there is an abrupt jump close to $|B|=1.6$~T at which all resistance features shift towards more positive $V_{gr}$, indicating a sudden transfer of electrons out of the graphene. To interpret this, we consider a simple model in which the CrI$_3$ is a bilayer in order to investigate the origin of this effect, justified by the expectation that the graphene couples most strongly to the nearest few layers of the CrI$_3$. Ab initio calculations show that the energy of the CrI$_3$ conduction band depends on its interlayer magnetic ordering, shifting to lower energy as the material undergoes a transition from interlayer AFM to FM (Fig.~\ref{fig:4}b). This transition likely also reduces the energy levels of defect states, and results in additional electrons tunneling into the CrI$_3$ (see band schematics in Fig.~\ref{fig:4}b). This abrupt jump is absent in Region I, where the states in the CrI$_3$ remain too high to be occupied (e.g., see Fig.~\ref{fig:2}b). This effect demonstrates that the graphene resistivity is highly sensitive to the interlayer magnetic ordering of the CrI$_3$. We find that the critical field becomes asymmetric with the sweeping direction of the magnetic field for thicker CrI$_3$ substrates (Supplementary Information Fig.~\ref{fig:CrI3_collective}), indicating that the graphene is only sensitive to the magnetic ordering of the nearest few CrI$_3$ layers.

The anomalous Hall effect (AHE) is anticipated in graphene endowed with both a magnetic exchange field and Rasbha spin orbit coupling~\cite{Qiao2010PRB,Qiao2014PRL,Zhang2015PRB,Zhang2018PRB,Hog2020PRB}, owing to the formation of an inverted gap at the Dirac point and associated Berry curvature at the band edges. We search for the AHE by measuring $\rho_{xy}$ as the magnetic field is swept back and forth at different values of $V_{gr}$ (Figure~\ref{fig:4}c). We see hysteretic loops surrounding $|B| \approx 1.6$~T owing to the AFM/FM transition in the CrI$_3$, but do not observe hysteresis at $B=0$. For measurements acquired nearby the Dirac point, we observe nonlinear $\rho_{xy}(B)$ surrounding $B=0$ reminiscent of a weak AHE. However, we also notice that the Dirac point drifts with magnetic field in the Landau fans shown in Fig.~\ref{fig:2}a and Fig.~\ref{fig:4}a. Although we do not understand the origin of this effect, and further find that it is highly sample dependent (e.g., see Supplementary Fig.~\ref{fig:10L_CrI3}f), its presence here implies that the charge transfer between the graphene and CrI$_3$ changes continuously with the magnetic field. As a result, the charge carrier density in graphene also changes with $B$ at fixed gate voltage, potentially driving a nonlinearity in the observed Hall effect that is completely unrelated to the usual AHE mechanism.

Careful analysis of the Landau fans can provide further insights into the strength of the magnetic proximity coupling in graphene. As detailed earlier, we observe a four-fold degeneracy in nearly all of the IQH states in our sample (with only weak signatures of symmetry-breaking at high $B$), indicating preserved spin and valley degeneracy. The absence of symmetry-broken IQH states is expected given the modest graphene mobility of $\sim5000$~cm$^2$/Vs (see Supplementary Information Fig.~\ref{fig:mobility}), presumably resulting from scattering due to defects in the CrI$_3$ substrate. However, the absence of detectable Landau-level splitting also sets an upper bound on the magnitude of the magnetic exchange coupling, which is expected to act as a Zeeman term that lifts the spin degeneracy at zero field. The magnetic exchange coupling must therefore be less than the smallest resolvable Landau level gap in our measurements. We estimate this to be $\sim$25~meV from the fact that the $\nu=-2$ state fully develops at $B\approx0.5$~T (Fig.~\ref{fig:4}c), following the expectation that the corresponding cyclotron gap is $\Delta E_{-2}=v_F\sqrt{2 \hbar e B}$, where $v_F=10^6$~m/s is the presumed Fermi velocity of graphene and $\hbar$ is Planck's reduced constant. This is in tension with theoretical predictions of magnetic exchange couplings ranging from tens of meV to as large as $\sim$120~meV in graphene on CrI$_3$~\cite{Holmes2020,Farooq2019,Zhang2018PRBCrI3}. Furthermore, unambiguous evidence for a proximity exchange field has been observed in optical spectroscopy measurements of monolayer WSe$_2$ on a CrI$_3$ substrate~\cite{Zhong2017SciAdv,Zhong2020NatNano}. The apparent unexpectedly small exchange coupling for graphene on CrI$_3$ may be intrinsic, but it may also be degraded by disorder in the CrI$_3$ for reasons that are not clear at present. Progress towards a high-temperature quantum anomalous Hall effect in proximitized graphene will likely require a reduction of the defect concentration in the CrI$_3$ crystal, or the discovery of more favorable magnetic insulator substrates.

\section*{Methods} 

\textbf{Device fabrication.} Samples are assembled using a dry-transfer technique with a polycarbonate (PC)/polydimethyl siloxane (PDMS) stamp~\cite{Wang2013}. The CrX$_3$ must remain in an inert environment during the entire vdW assembly process to avoid crystal degradation. In order to achieve this, we first shape a flake of monolayer graphene into a Hall bar geometry using a polymer-free anodic oxidation lithography technique with an atomic force microscope tip~\cite{Li2018AON}. The vdW heterostructure consists of a graphene/CrX$_3$ interface encapsulated between flakes of boron nitride. Additional graphite flakes enclad some samples to act as gates. The sample is deposited onto a Si/SiO$_2$ wafer after assembly by melting the PC film at 180$^{\circ}$C. The entire vdW heterostructure assembly is performed inside a glovebox filled with argon. It is then removed from the glovebox, and the PC film is dissolved in chloroform. The CrX$_3$ crystals are protected from degradation as long as they remain fully encapsulated by BN flakes. We use standard electron beam lithography, CHF$_3$/O$_2$ plasma etching, and metal deposition techniques (Cr/Au) in order to electrically contact the graphene Hall bar in regions far from the CrX$_3$ flake. Supplementary Fig.~\ref{fig:device_fab} illustrates the fabrication procedure.

\textbf{Transport measurements.} Transport measurements were performed either in a Cryomagnetics variable temperature insert or a Quantum Design DynaCool PPMS, and were conducted in a four-terminal geometry with a.c. current excitation of 10-100 nA using standard lock-in techniques at a frequency of 13.3 Hz. In some cases, a gate voltage was applied to the Si gate in order to dope the region of the graphene contacts overhanging the graphite back gate to a high charge carrier density and reduce the contact resistance. The graphene resistivity, $\rho$, is calculated from the measured resistance, $R$, as  $\rho=(w/l) R$, where $w$ is the Hall bar width and $l$ is the distance between the centers of the voltage probes of the graphene Hall bar. Unless otherwise stated, all measurements were performed at a base temperature of $T=1.5-2$~K.

\textbf{Reproducibility of the transport in graphene/CrI$_3$.} We have characterized two additional graphene/CrI$_3$ devices with CrI$_3$ thicknesses of 7 and 10 layers (Supplementary Information Figs.~\ref{fig:10L_CrI3}-\ref{fig:7L_CrI3}), as well as an additional 3 layer device with a monolayer of WSe$_2$ sandwiched between the graphene and CrI$_3$ (Supplementary Information Figs.~\ref{fig:CrI3_WSe2}-\ref{fig:WSe2_Hall}). Transport in all devices closely resembles that described in the device presented in the main text. 

\textbf{Transport in graphene on CrBr$_3$ and CrCl$_3$.} In graphene on CrBr$_3$ and CrCl$_3$, the transport properties over the entire accessible gate voltage range closely resemble those of Region III for graphene on CrI$_3$ (see Supplementary Figs.~\ref{fig:CrBr3_device}-\ref{fig:CrCl3_thin}). We see negative differential capacitance upon tuning $V_{mag}$ over the entire accessible gate range, as well as hysteretic graphene transport. The differences among the CrX$_3$ substrates arise as a consequence of the larger electron affinities of the CrBr$_3$ and CrCl$_3$, resulting in larger modulation doping of the graphene that cannot be suppressed by gating (i.e., we are unable to access the equivalent of Region I).

\textbf{Temperature-dependent transport features at the onset of magnetic ordering.} Supplementary Information Fig.~\ref{fig:temperature} shows temperature-dependent transport measurements of graphene on CrI$_3$, CrBr$_3$, and CrCl$_3$ across the critical temperature for magnetic ordering of each material. The graphene resistivity clearly changes within a few kelvin of the anticipated magnetic ordering temperature (dashed lines) for the case of CrI$_3$, providing further evidence for the magnetic proximity effect. Although we do not understand the detailed origin of these features, they likely arise from a modification of the charge transfer as the CrI$_3$ magnetically orders owing to a small energy shift of the bands. Temperature-dependent resistivity changes are also observed for CrBr$_3$ and CrCl$_3$, however they arise over a much broader range of temperature than for CrI$_3$, and their connection to the onset of magnetic ordering is less obvious. 

\section*{Acknowledgments}
This work was supported as part of Programmable Quantum Materials, an Energy Frontier Research Center funded by the U.S. Department of Energy (DOE), Office of Science, Basic Energy Sciences (BES), under award DE-SC0019443. M.Y., X.X., and J.-H.C. acknowledge support from the State of Washington funded Clean Energy Institute. This work made use of shared fabrication facilities provided by NSF MRSEC 1719797. The material synthesis at UW is partially supported by the Gordon and Betty Moore Foundation’s EPiQS Initiative, Grant GBMF6759 to J.-H.C. K.W. and T.T. acknowledge support from the Elemental Strategy Initiative conducted by the MEXT, Japan (Grant Number JPMXP0112101001) and JSPS KAKENHI (Grant Numbers 19H05790, 20H00354 and 21H05233). M.M.’s crystal synthesis effort at ORNL was supported by the US Department of Energy, Office of Science, Basic Energy Sciences, Materials Sciences and Engineering Division. \\

\section*{Author contributions}
C.-C.T. and T.S. fabricated the devices, with assistance from J.S. C.-C.T. and T.S. performed the measurements. Q.J., Z.L., and J.-H.C. provided the bulk CrI$_3$ and CrBr$_3$ crystals. M.M. provided the bulk CrCl$_3$ crystals. K.W. and T.T. provided the bulk BN crystals. C.W. and D.X. calculated the CrI$_3$ band structure. C.-C.T. and T.S. analyzed the data under the supervision of D.H.C., X.X. and M.Y.

\section*{Data Availability}
Source data are available for this paper. All other data that support the plots within this paper and other findings of this study are available from the corresponding author upon reasonable request.

\section*{Competing interests}
The authors declare no competing interests.

\section*{Additional Information}
Correspondence and requests for materials should be addressed to M.Y.

\section*{Supplementary Information}
Supplementary Sections S1-S10 and Figs. S1-S18.

\bibliographystyle{naturemag}
\bibliography{references}

\clearpage

\renewcommand{\thefigure}{S\arabic{figure}}
\renewcommand{\thesection}{S\arabic{section}}
\renewcommand{\thesubsection}{S\arabic{subsection}}
\renewcommand{\theequation}{S\arabic{equation}}
\renewcommand{\thetable}{S\arabic{table}}
\setcounter{figure}{0} 
\setcounter{equation}{0}
\appendix 

\onecolumngrid

\section*{Supplementary Information}

\textbf{S1. Summary of the graphene/CrX$_3$ devices}\\

\begin{figure*}[h]
\includegraphics[width=6.9 in]{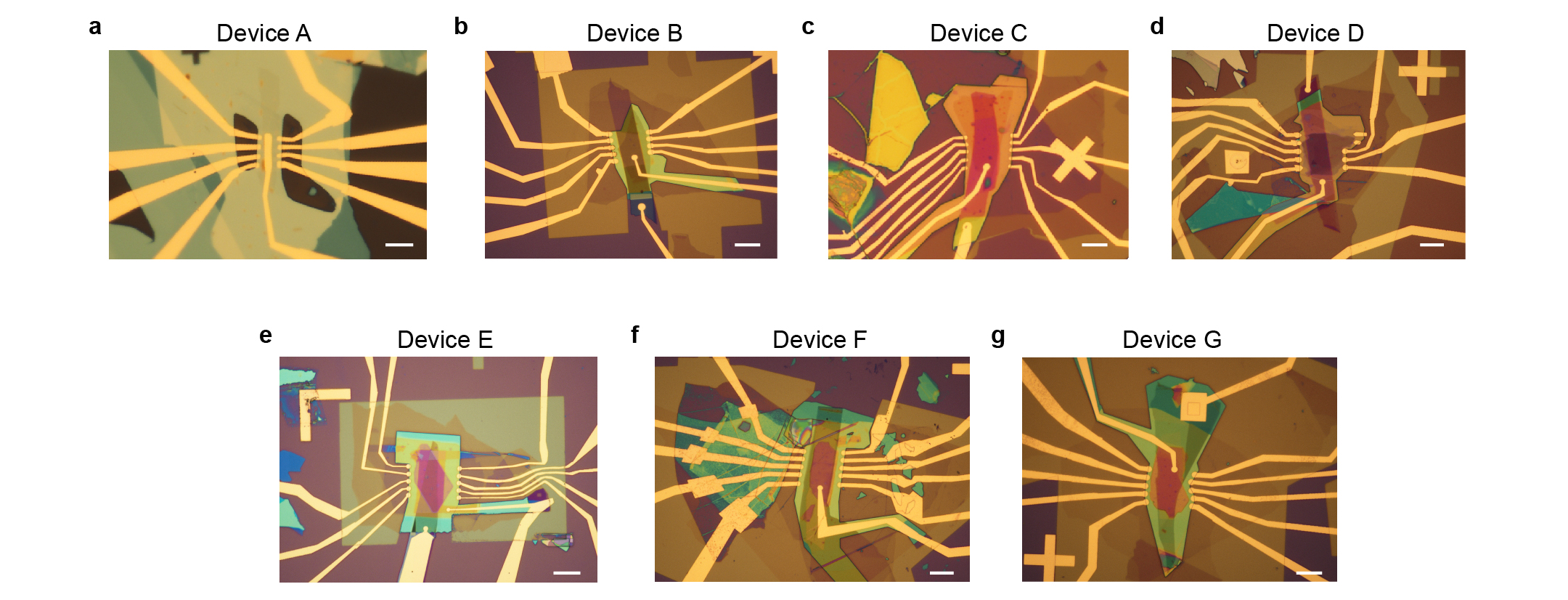} 
\caption{\textbf{Optical micrographs of the seven devices in this study.}
All scale bars are 10~$\mu$m.
}
\label{fig:device_images}
\end{figure*}

\begin{figure*}[h]
\includegraphics[width=6.9 in]{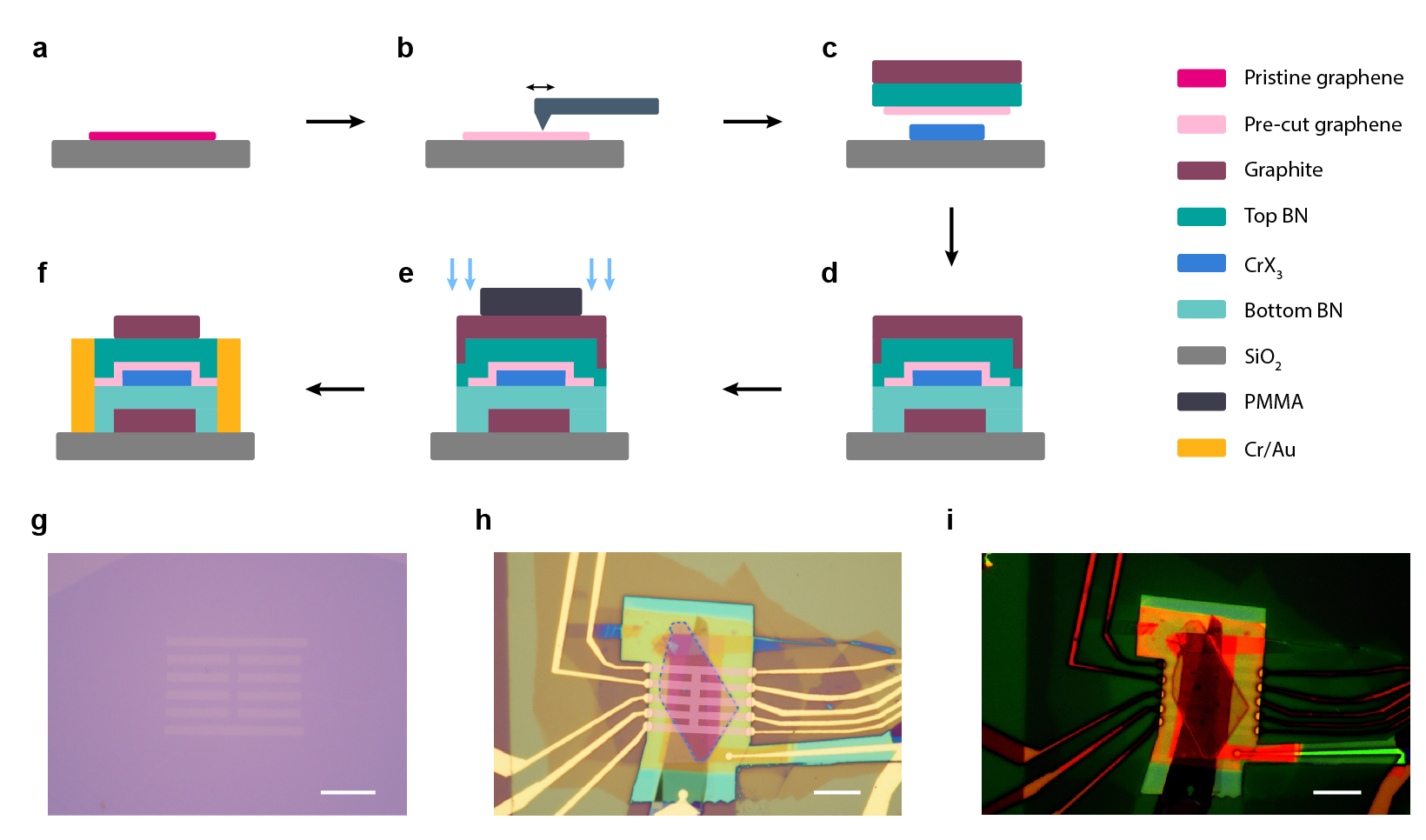} 
\caption{\textbf{Device fabrication procedure.}
\textbf{a}, Flakes of monolayer graphene, graphite, BN, and CrX$_3$ are isolated by mechanical exfoliation. \textbf{b}, The graphene is cut into a Hall bar geometry using a polymer-free anodic oxidation lithography technique with an atomic microscope tip. \textbf{c}, The vdW layers are assembled using conventional dry transfer techniques with a PC stamp. The entire transfer process is conducted inside an inert gas glovebox (filled with Ar), with oxygen and water levels below 1 ppm. \textbf{d}, The vdW heterostructure is deposited onto a clean Si/SiO$_2$ chip and removed from the glovebox. The CrX$_3$ is protected from degredation by the encapsulating top and bottom BN crystals. \textbf{e}, The device is etched with CHF$_3$ and O$_2$ plasma using a protective PMMA mask. The CrX$_3$ remains completely encapsulated in BN after the etch procedure, however, the arms of the graphene Hall bar are exposed. \textbf{f}, Cr/Au is evaporated to form electrical contacts to the graphene Hall bar and the graphite gates. 
\textbf{g-i}, Optical micrographs of a flake of monolayer graphene cut by an AFM tip to form an internal Hall bar geometry (\textbf{g}), a representative completed device (Device E) (\textbf{h}), and the same device with digital color filtration to make the graphene Hall bar arms visible (\textbf{i}). The pink Hall bar overlay in \textbf{h} indicates the region of the pre-cut graphene, and blue dashed curve outlines the CrBr$_3$ flake. All scale bars are 10~$\mu$m.
}
\label{fig:device_fab}
\end{figure*}

We study six devices of graphene on various CrX$_3$ substrates, and one device with a monolayer WSe$_2$ spacer. Table~\ref{tab:summary} summarizes the details of these devices. Data in the main text is acquired from Device A. Figure~\ref{fig:device_images} shows optical micrographs of all seven devices. Device A has a silicon back gate and gold top gate. All other devices have graphite top and bottom gates. Figure~\ref{fig:device_fab} illustrates the fabrication procedure for our devices (see Methods for additional discussion of the device fabrication).

\begin{table*}[h]
\centering
\begin{tabularx}{0.97\textwidth} { 
  | >{\centering\arraybackslash}X 
  | >{\centering\arraybackslash}X 
  | >{\centering\arraybackslash}X
  | >{\centering\arraybackslash}X  }
\hline
Device & Material & CrX$_3$ Thickness \\ 
\hhline{|=|=|=|}
A & CrI$_3$ & 3 layers \\ 
\hline
B & CrI$_3$ & 7 layers \\ 
\hline
C & CrI$_3$ & 10 layers \\ 
\hline
D & WSe$_2$/CrI$_3$ & 1/3 layers \\ 
\hline
E & CrBr$_3$ & 28 nm \\ 
\hline
F & CrCl$_3$ & 40 nm \\ 
\hline
G & CrCl$_3$ & 80 nm \\ 
\hline
\end{tabularx}
\caption{Summary of the devices reported in our study.}
\label{tab:summary}
\end{table*}

\clearpage

\textbf{S2. Determination of the carrier type and Dirac point trajectory in Device A}\\

\begin{figure*}[h]
\includegraphics[width=6.9 in]{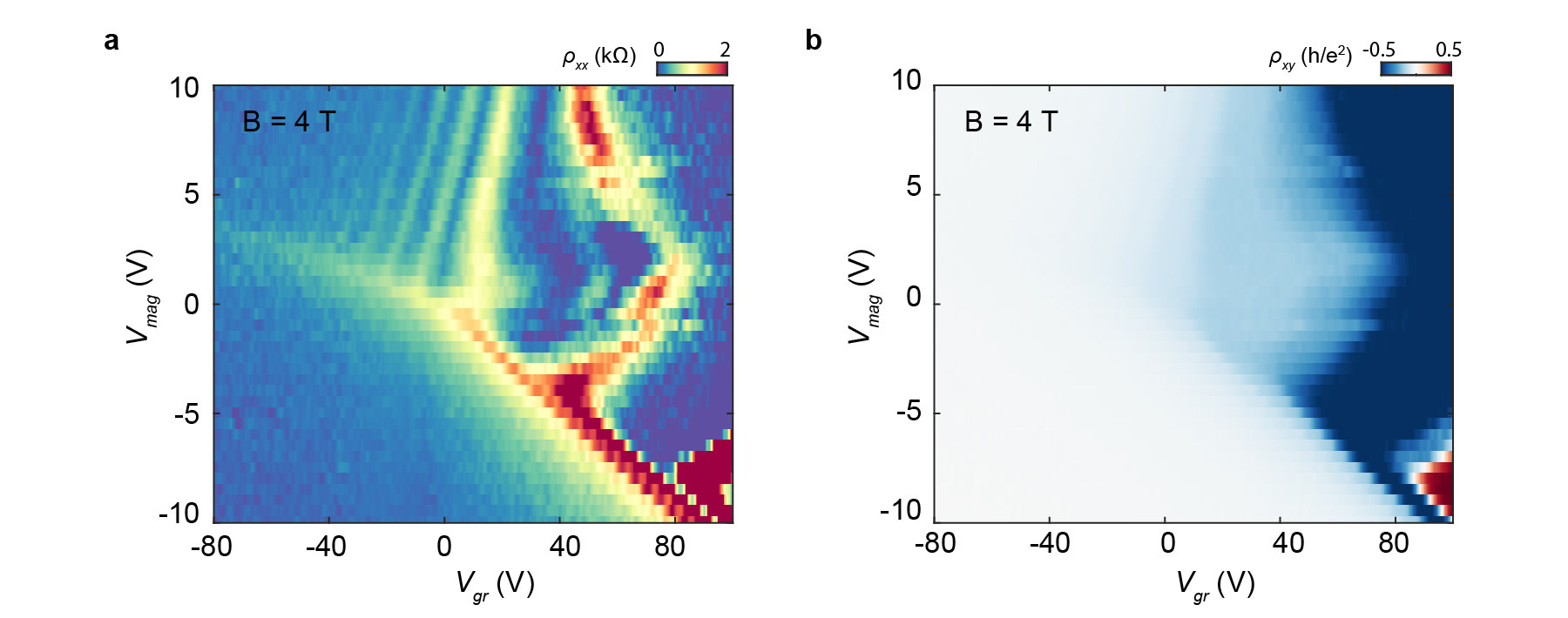} \caption{\textbf{Transport measurements of Device A at $B=4$~T.}
\textbf{a-b}, Maps of $\rho_{xx}$ (\textbf{a}) and $\rho_{xy}$ (\textbf{b}) acquired by sweeping both gates at $B=4$~T. The sign of the Hall effect is negative over the entire map except for a small region in the bottom right, indicating that the graphene is hole-doped. The blue dashed curve in Fig.~1d of the main text corresponds to the condition of $\rho_{xy}=0$ in (\textbf{b}).
}
\label{fig:DeviceA_DP4T}
\end{figure*}

\clearpage

\clearpage

\textbf{S3. Transport in Device A acquired in the reverse gate sweeping condition}\\

\begin{figure*}[h]
\includegraphics[width=5 in]{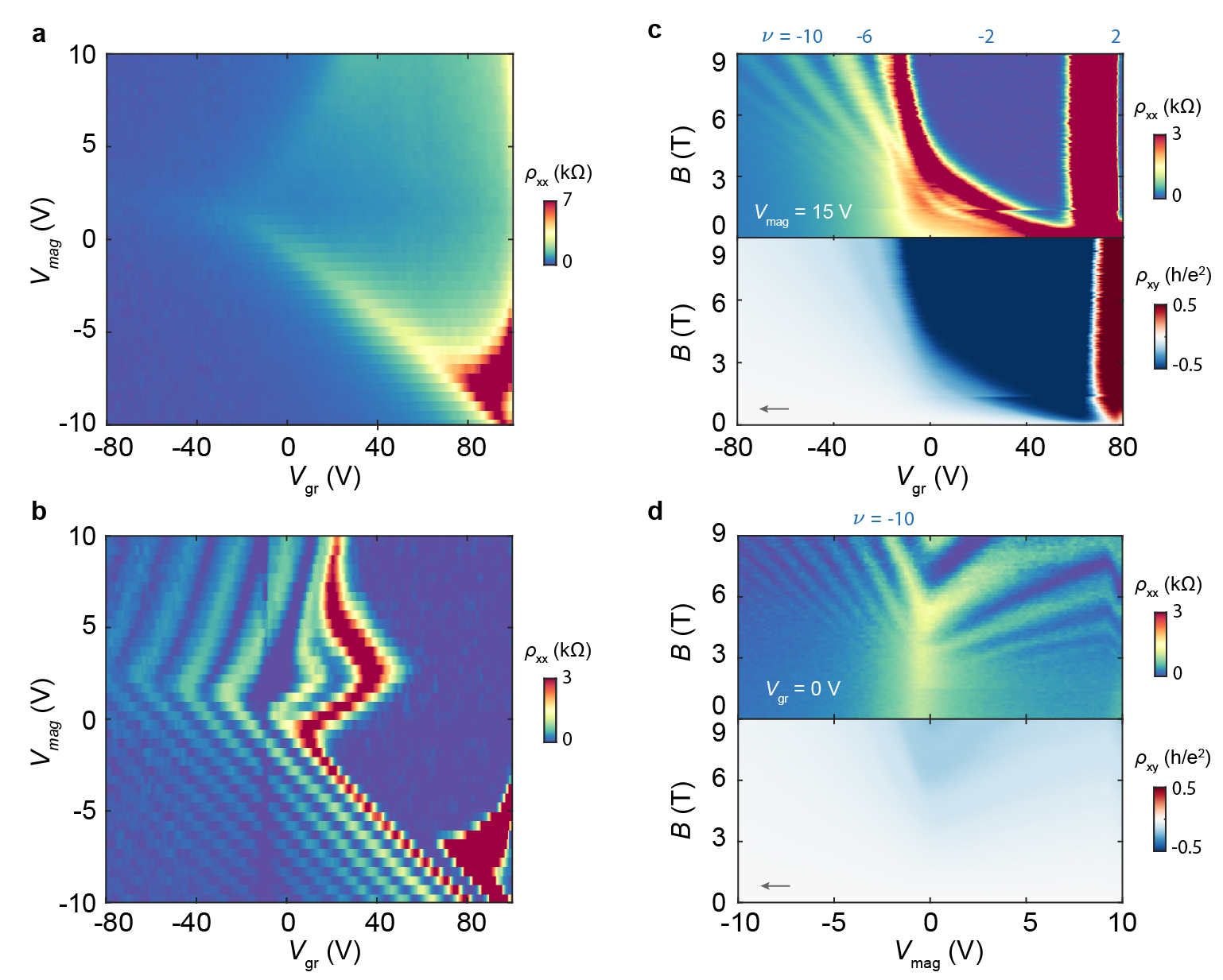} 
\caption{\textbf{Transport measurements of Device A with the reverse gate sweeping direction.}
\textbf{a-b}, Maps of $\rho_{xx}$ acquired by sweeping both gates at $B=0$~T (\textbf{a}) and $B=9$~T (\textbf{b}). In contrast to Fig.~1d and Fig.~3a of the main text, these maps are acquired by sweeping the fast axis ($V_{gr}$) from positive to negative. 
\textbf{c-d}, Landau fan diagrams acquired by sweeping $V_{gr}$ from negative to positive with $V_{mag}=15$~V (\textbf{c}), and $V_{mag}$ with $V_{gr}=0$ (\textbf{d}), respectively. These are otherwise the same as the measurements shown in Figs.~2a and c of the main text.
}
\label{fig:DeviceA_reverse}
\end{figure*}

All transport data for Device A in the main text is acquired by sweeping the fast-axis gate from negative to positive values. Figure~\ref{fig:DeviceA_reverse} shows comparable measurements acquired by instead sweeping the fast-axis gate from positive to negative. Transport differs primarily in Regions II and III, as evidenced by the hysteresis maps in Figs.~1e and 3b (inset) of the main text. 

\clearpage

\textbf{S4. Transport characterization of additional graphene/CrI$_3$ devices}\\

\begin{figure*}[h]
\includegraphics[width=6.9 in]{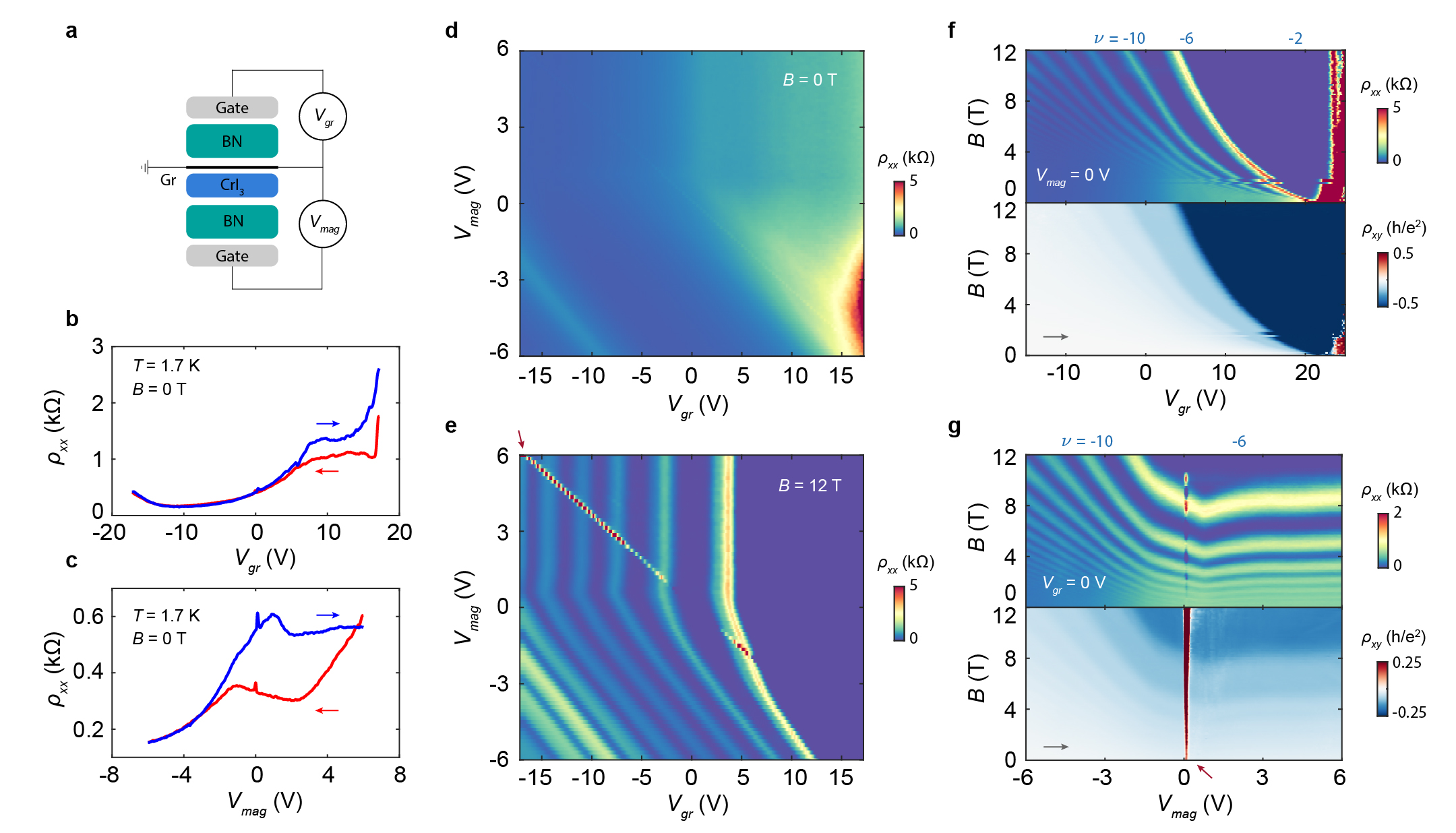} 
\caption{\textbf{Transport measurements of graphene on a 10 layer CrI$_3$ (Device C).}
\textbf{a}, Cartoon schematic of the device structure. \textbf{b}, $\rho_{xx}$ as $V_{gr}$ is swept back and forth with $V_{mag}=0$~V. \textbf{c}, $\rho_{xx}$ as $V_{mag}$ is swept back and forth with $V_{gr}=0$~V. 
\textbf{d-e}, Maps of $\rho_{xx}$ acquired by sweeping both gates at $B=0$~T (\textbf{d}) and $B=12$~T (\textbf{e}), respectively. 
\textbf{f-g}, Landau fan diagrams acquired by sweeping $V_{gr}$ (\textbf{f}) and $V_{mag}$ (\textbf{g}), respectively, from negative to positive with the other gate held at ground. Features denoted by red arrows in \textbf{e-g} are artifacts arising due to insulating behavior at the contacts.
}
\label{fig:10L_CrI3}
\end{figure*}

We have studied two graphene/CrI$_3$ devices (Devices B and C) in addition to the one reported in the main text (Device A). These two devices have thicker CrI$_3$ substrates of 7 and 10 layers, respectively. Figure~\ref{fig:10L_CrI3}a shows a schematic of Device C, which has graphite top and bottom gates. Figures~\ref{fig:10L_CrI3}b-c show the device resistivity acquired by sweeping one gate back and forth with the other grounded, and Fig.~\ref{fig:10L_CrI3}d shows the resistivity map acquired by sweeping both gates, analogous to the measurements of Figs.~1b-d of the main text for Device A. The basic transport phenomenology is nearly identical between the two devices, including the gate-tunable modulation doping, the bent trajectory of the Dirac point, anomalous resistive peaks and plateaus in the hole-doped regime, and hysteresis confined only to certain portions of the phase diagram. We note that this device appears to be aligned with the encapsulating boron nitride and manifests a weak secondary Dirac point at large hole doping (diagonal feature in the bottom left of Fig.~\ref{fig:10L_CrI3}d), however, this does not appear to have a meaningful impact on the overall transport properties arising from the graphene/CrI$_3$ interface. Figure~\ref{fig:10L_CrI3}e shows a similar resistivity map at $B=12$~T, and Figs.~\ref{fig:10L_CrI3}f-g show Landau fan diagrams of the longitudinal (top) and Hall (bottom) resistivities acquired by sweeping $V_{gr}$ and $V_{mag}$, respectively, with the other gate held at ground. Again, we see that this device reproduces the salient high-field features of Device A reported in the main text, including a highly extended $\nu=-2$ plateau, nonlinear trajectories of the IQH states, signatures of negative compressibility when tuning $V_{mag}$ in the equivalent of Region III, and an abrupt jump in the modulation doping around the magnetic field at which the CrI$_3$ becomes a layer ferromagnet.

\begin{figure*}[h]
\includegraphics[width=5.5 in]{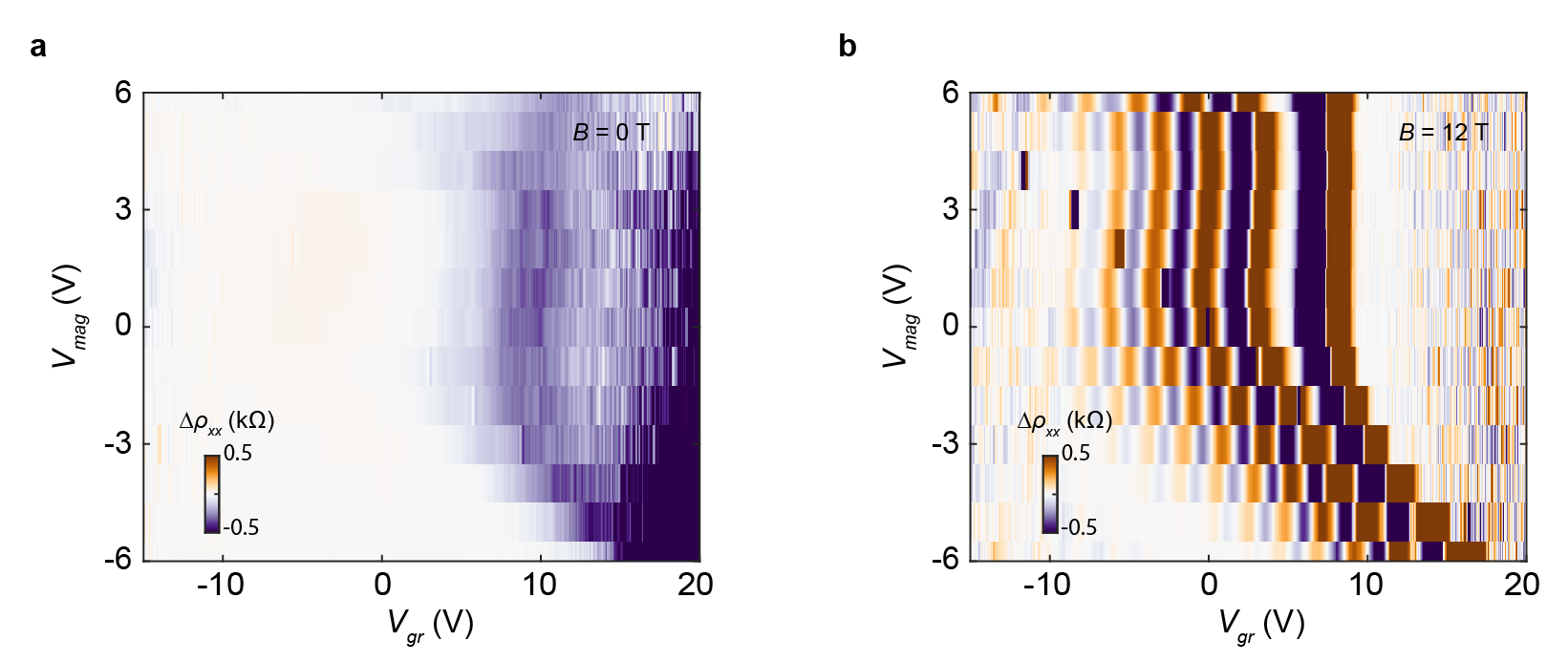} 
\caption{\textbf{Map of the transport hysteresis in Device C.}
\textbf{a-b}, $\Delta \rho_{xx}$ acquired by taking the difference between the resistivity upon sweeping $V_{gr}$ forward and backward at $B=0$~T (\textbf{a}) and $B=12$~T (\textbf{b}).
}
\label{fig:DeviceC_hysteresis}
\end{figure*}

\begin{figure*}[h]
\includegraphics[width=6.5 in]{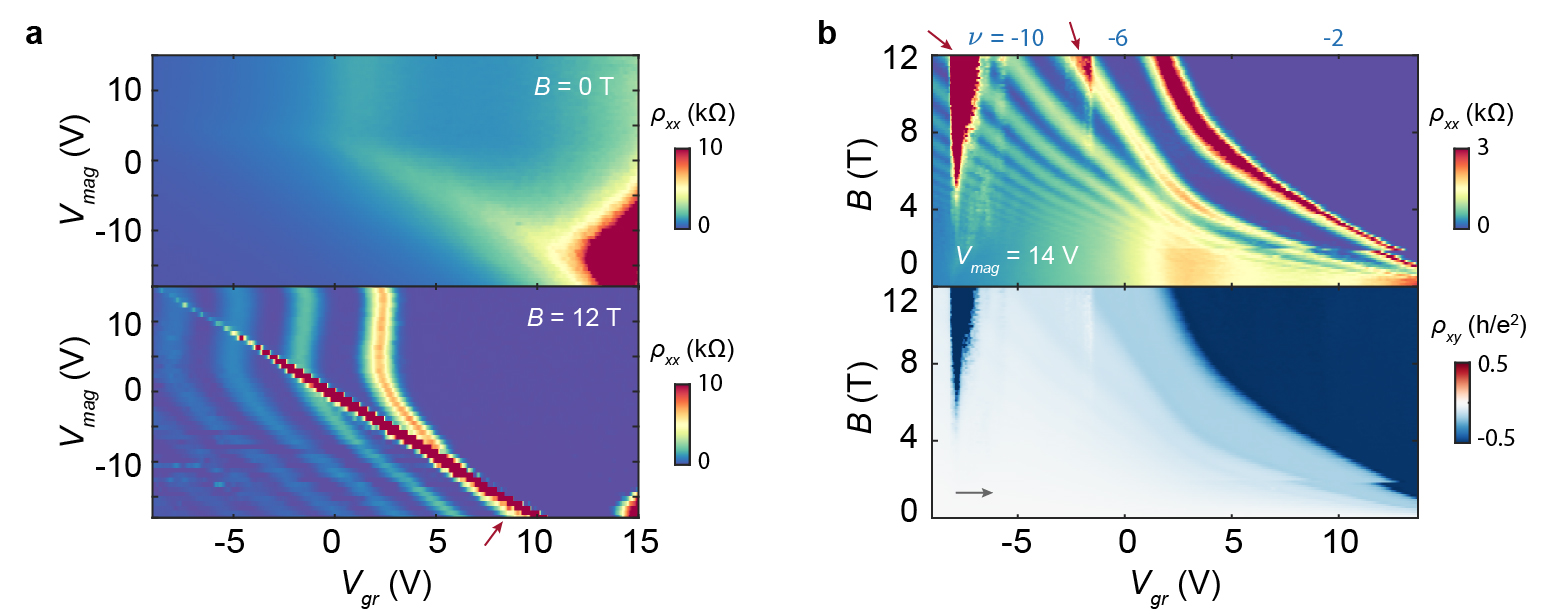} 
\caption{\textbf{Transport measurements of graphene on a 7 layer CrI$_3$ (Device B).}
\textbf{a}, Maps of $\rho_{xx}$ acquired by sweeping both gates at  $B=0$~T (top) and $B=12$~T (bottom). \textbf{b}, Landau fan diagram acquired by sweeping  $V_{gr}$ from negative to positive values with $V_{mag}=14$~V. Features denoted by red arrows are artifacts arising due to insulating behavior at the contacts.
}
\label{fig:7L_CrI3}
\end{figure*}

Figures~\ref{fig:DeviceC_hysteresis}a-b show the hysteresis maps of this device acquired at $B=0$ and 12~T, respectively. As in Device A, the combination of these measurements reveals hysteretic behavior everywhere except the bottom left portion of the map (Region I). Device B also behaves quite similarly overall (Fig.~\ref{fig:7L_CrI3}), indicating excellent reproducibility of the transport properties of graphene/CrI$_3$ heterostructures.

\clearpage

\textbf{S5. Transport characterization of graphene on CrBr$_3$ and CrCl$_3$}\\

\begin{figure*}[h]
\includegraphics[width=6.9 in]{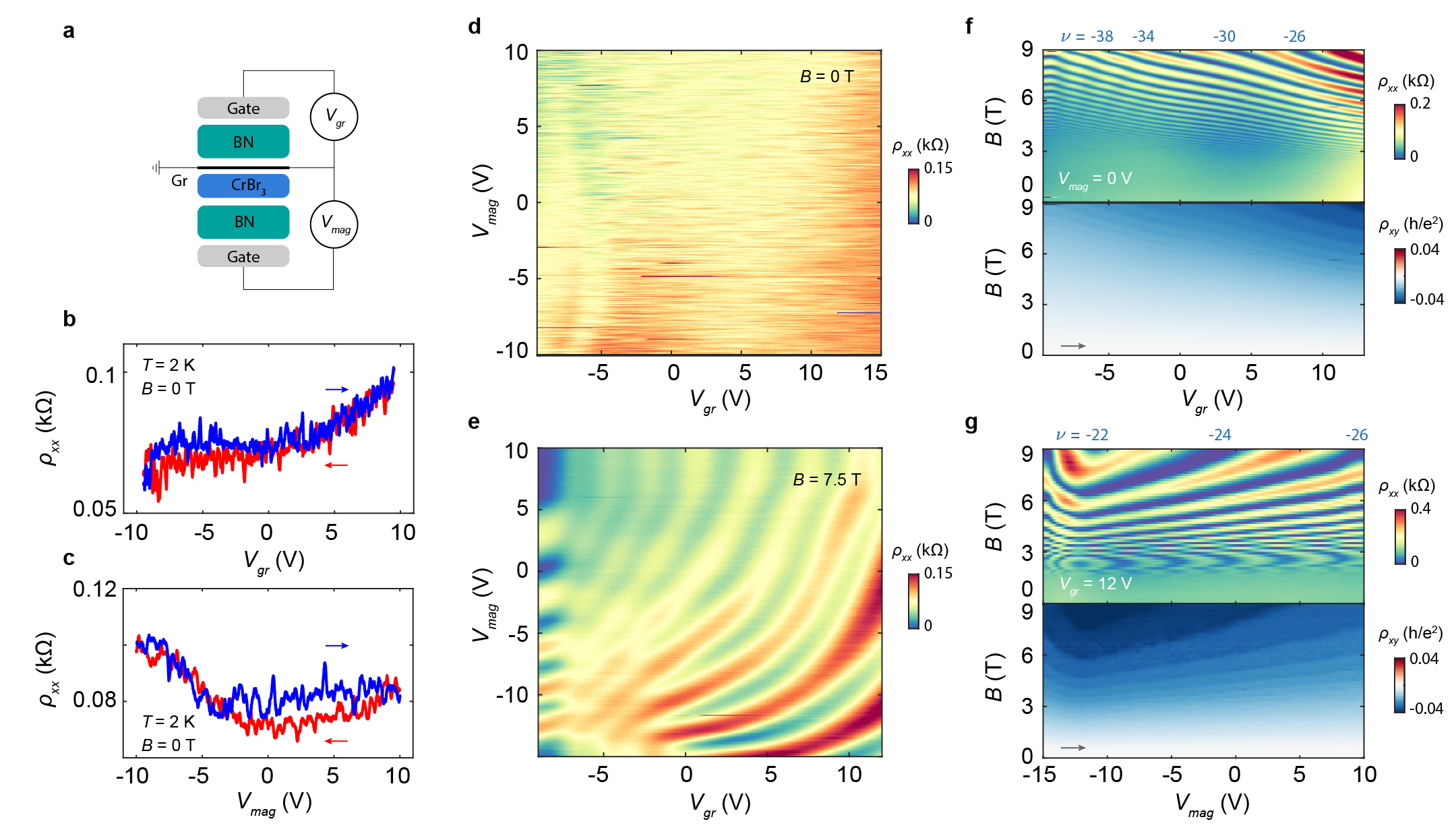} 
\caption{\textbf{Transport measurements of graphene on a 28 nm CrBr$_3$ (Device E).}
\textbf{a}, Cartoon schematic of the device structure. \textbf{b}, $\rho_{xx}$ as $V_{gr}$ is swept back and forth
with $V_{mag}=0$~V. \textbf{c}, $\rho_{xx}$ as $V_{mag}$ is swept back and forth with $V_{gr}=0$~V. \textbf{d-e}, Maps of $\rho_{xx}$ acquired by sweeping both gates at $B=0$~T (\textbf{d}) and $B=7.5$~T (\textbf{e}), respectively. 
\textbf{f-g}, Landau fan diagrams acquired by sweeping $V_{gr}$ from negative to positive with $V_{mag}=0$ (\textbf{f}), and $V_{mag}$ with $V_{gr}=12$~V (\textbf{g}), respectively.
}
\label{fig:CrBr3_device}
\end{figure*}

We have performed transport characterization similar to that described in the main text for graphene/CrBr$_3$ and graphene/CrCl$_3$ heterostructures. CrBr$_3$ is both an intra- and inter-layer ferromagnet at low temperature. We construct a dual graphite-gated device with a graphene/CrBr$_3$ interface (Fig.~\ref{fig:CrBr3_device}a) and measure the graphene resistivity as a function of each gate (Figs.~\ref{fig:CrBr3_device}b-c). Upon sweeping both gates, we find that the graphene resistivity is very low over the entire accessible range of doping (Fig.~\ref{fig:CrBr3_device}d), with much smaller gate-induced changes in resistivity than in graphene/CrI$_3$. In a similar measurement at $B=7.5$~T, we see IQH states that exhibit highly unusual trajectories as a function of gating. These states move nonlinearly with gate voltage, and disperse in the opposite direction than anticipated upon gating with $V_{mag}$ indicative of an apparent negative compressibility. 

These observations are corroborated by measurements of the Landau fan diagrams. In the fan acquired by tuning $V_{gr}$ at fixed $V_{mag}=0$ (Fig.~\ref{fig:CrBr3_device}f), we observe magnetotransport that is mostly consistent with typical highly-doped graphene. The four-fold degenerate IQH states disperse nearly linearly with $B$, projecting to a very positive $V_{gr}$ indicating large modulation hole doping. Corresponding measurements of $\rho_{xy}$ further corroborate the large filling factors of the IQH states and their hole-type doping. In contrast, the Landau fan acquired by tuning $V_{mag}$ at fixed $V_{gr}=12$~V is highly atypical, with IQH states dispersing away from the apparent Dirac point over the majority of the accessible gate range (Fig.~\ref{fig:CrBr3_device}g). The upturn of these states at very negative $V_{mag}$ is a consequence of hysteresis in the system. These oppositely dispersing IQH states indicate the negative differential capacitance of the system upon tuning $V_{mag}$ over the entire accessible gate range of the device. These observations are phenomenologically consistent with Region III of the graphene/CrI$_3$ phase diagram, and indicate that the electron affinity of CrBr$_3$ is larger than CrI$_3$. In this case, the lowest electron band of CrBr$_3$ resides below the Dirac point of graphene, resulting in modulation doping that cannot be suppressed with gating. 

\begin{figure*}[h]
\includegraphics[width=3 in]{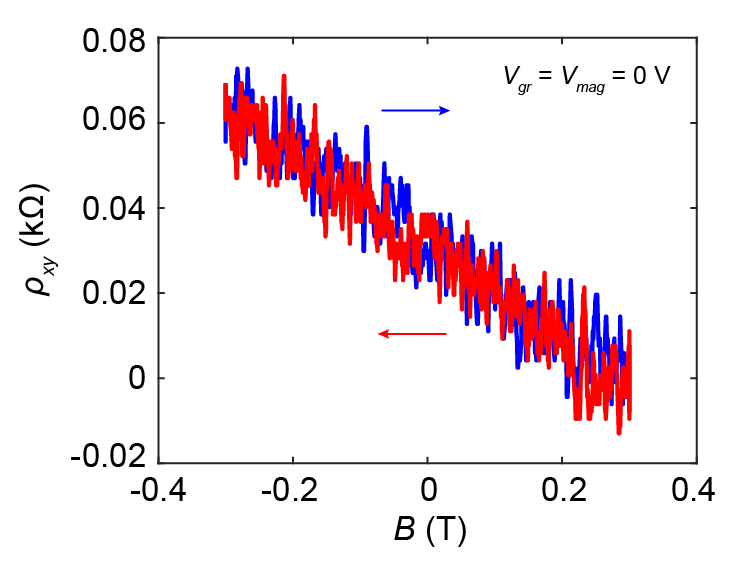} 
\caption{\textbf{Measurement of the Hall resistance in Device E.}
$\rho_{xy}$ measured as $B$ is swept back and forth with $V_{gr}=V_{mag}=0$~V.
}
\label{fig:CrBr3_Hall}
\end{figure*}

We do not find any signs of a magnetic exchange field in the graphene, very similar to the case of graphene/CrI$_3$. We observe an ordinary (linear) Hall effect over the entire accessible gate range (see Fig.~\ref{fig:CrBr3_Hall} for a representative example), although this is expected even in the case of a sizable magnetic exchange field since the Berry curvature is likely concentrated close to the Dirac point. However, we do not observe any signatures of degeneracy lifting in the main-sequence IQH states. We follow a similar analysis of the quantum Hall gap as described in the main text to estimate the approximate upper bound strength of the magnetic exchange field as $\sim$15~meV, which is similar to that of graphene/CrI$_3$. In contrast, calculations suggest exchange couplings of ~$\sim$70~meV in graphene on CrBr$_3$~\cite{Behera2019CrBr3}. As in graphene/CrI$_3$, this rough upper bound estimate is much less than the theoretical expectation, and the reason for the discrepancy is not presently clear.

\begin{figure*}[h]
\includegraphics[width=6.9 in]{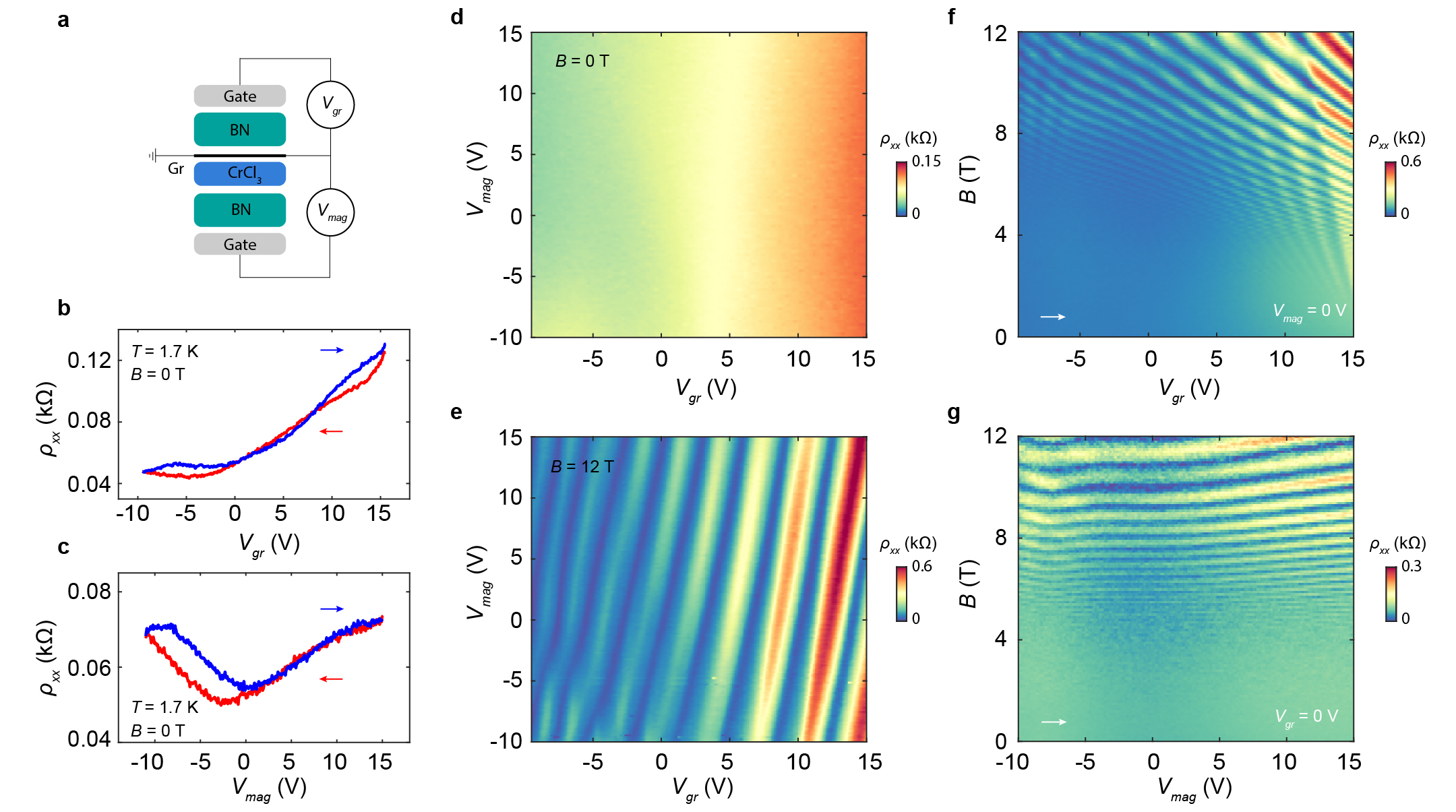} 
\caption{\textbf{Transport measurements of graphene on a 80 nm CrCl$_3$ (Device G).}
\textbf{a}, Cartoon schematic of the device structure. \textbf{b}, $\rho_{xx}$ as $V_{gr}$ is swept back and forth
with $V_{mag}=0$~V. \textbf{c}, $\rho_{xx}$ as $V_{mag}$ is swept back and forth with $V_{gr}=0$~V.
\textbf{d-e}, Maps of $\rho_{xx}$ acquired by sweeping both gates at $B=0$~T (\textbf{d}) and $B=12$~T (\textbf{e}), respectively. 
\textbf{f-g}, Landau fan diagrams acquired by sweeping $V_{gr}$ (\textbf{f}) and $V_{mag}$ (\textbf{g}), respectively, from negative to positive with the other gate held at ground.
}
\label{fig:CrCl3_thick}
\end{figure*}

\begin{figure*}[h]
\includegraphics[width=6.9 in]{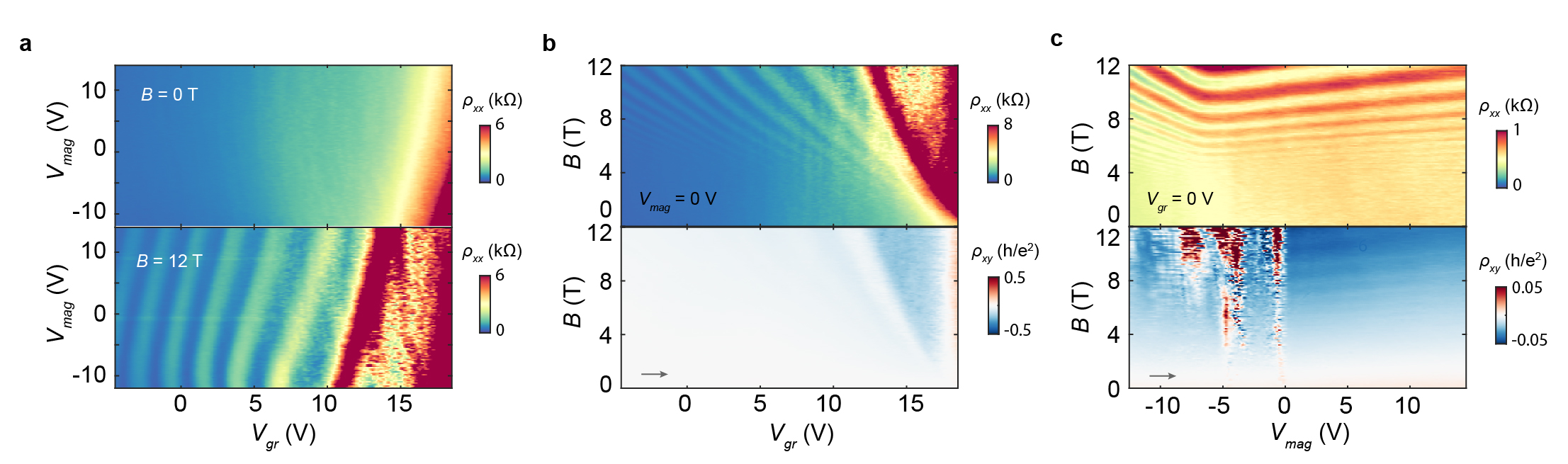} 
\caption{\textbf{Transport measurements of graphene on a 40 nm CrCl$_3$ (Device F).}
\textbf{a}, Maps of $\rho_{xx}$ acquired by sweeping both gates at  $B=0$~T (top) and $B=12$~T (bottom). \textbf{b}, Landau fan diagram acquired by sweeping  $V_{gr}$ from negative to positive values with $V_{mag}=0$~V. \textbf{c}, Landau fan diagram acquired by sweeping  $V_{mag}$ from negative to positive values with $V_{gr}=0$~V.
}
\label{fig:CrCl3_thin}
\end{figure*}

CrCl$_3$ is an intralayer ferromagnet and interlayer antiferromagnet. Spins in CrCl$_3$ are aligned in the 2D plane, rather than perpendicular as in CrI$_3$ and CrBr$_3$, although bulk CrCl$_3$ requires an out-of-plane field of only $\approx0.25$~T to fully polarize its spins out of the 2D plane~\cite{McGuire2017}. In spite of the different magnetic structure, we find that transport in graphene/CrCl$_3$ heterostructures is qualitatively consistent with CrBr$_3$. Figure~\ref{fig:CrCl3_thick}a shows the device structure for a sample of graphene on an 80 nm thick CrCl$_3$, and Figs.~\ref{fig:CrCl3_thick}b-g show transport measurements analogous to those presented for the graphene/CrBr$_3$ device in Fig.~\ref{fig:CrBr3_device}. Although the exact details differ, the basic properties look very similar. We find that the apparent modulation doping is slightly less in a second device with a 40 nm CrCl$_3$ (Fig.~\ref{fig:CrCl3_thin}, as we are just barely able to access the Dirac point by gating. At present, it is not clear what controls the precise amount of modulation doping in graphene/CrCl$_3$ devices. Despite this ambiguity, the majority of the accessible gate range in both devices phenomenologically corresponds to the behavior of Region III described in the main text.

\clearpage

\textbf{S6. Temperature-dependent transport of graphene/CrX$_3$}\\

\begin{figure*}[h]
\includegraphics[width=5 in]{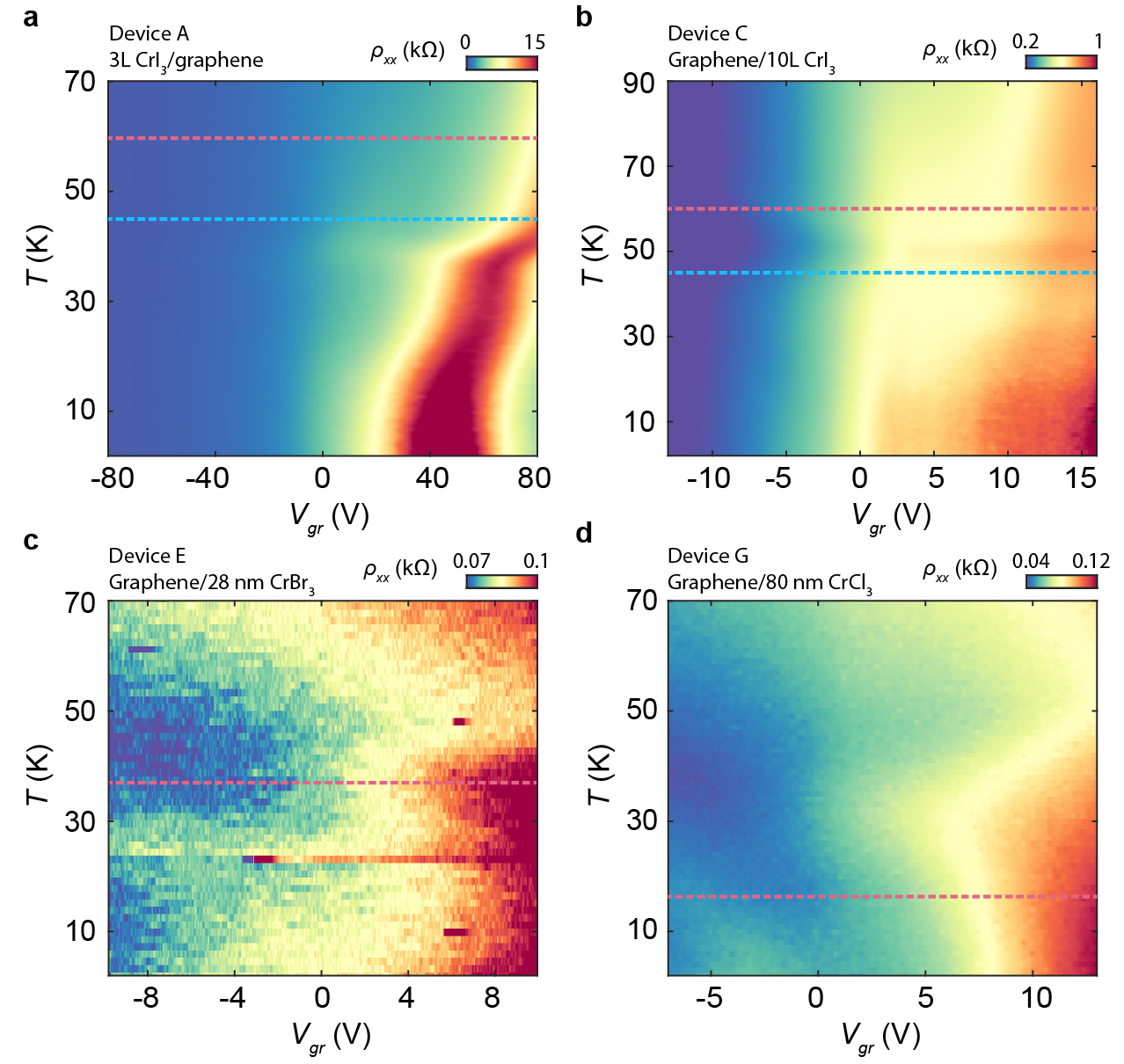} 
\caption{\textbf{Temperature dependence the graphene resistivity on various CrX$_3$ substrates.}
\textbf{a-b}, $\rho$(T) as a function of $V_{gr}$ in graphene/CrI$_3$ for Device A (\textbf{a}) and Device C (\textbf{b}). The blue (red) dashed line indicates the magnetic ordering temperature for monolayer (bulk) CrI$_3$.
\textbf{c-d}, Similar measurements for \textbf{c}, CrBr$_3$ (Device E) and \textbf{d}, CrCl$_3$ (Device G). The red dashed lines indicate the bulk magnetic ordering temperature of these materials. $V_{mag}$ is 15~V in \textbf{a}, 0~V in \textbf{b} and \textbf{d}, and 10~V in \textbf{c}.
}
\label{fig:temperature}
\end{figure*}

Figure~\ref{fig:temperature} shows the resistivity of graphene on CrI$_3$, CrBr$_3$, and CrCl$_3$ measured as a function of temperature and $V_{gr}$ at fixed $V_{mag}$. The overall resistivity drifts with temperature in ways that we do not fully understand, however, in the case of CrI$_3$ we see clear resistivity jumps within a few kelvin of the anticipated magnetic ordering temperature (denoted by the blue dashed line for the case of monolayer CrI$_3$ and the red dashed line for the case of bulk CrI$_3$~\cite{Huang2020review}). This indicates that the onset of magnetism in the CrI$_3$ impacts the graphene transport, likely by changing the magnitude of the modulation doping as the band edge shifts. Similar effects have been observed previously in graphene/RuCl$_3$ heterostructures~\cite{Zhou2019RuCl3}. The resistivity drifts over a much wider range of temperature for CrBr$_3$ and CrCl$_3$ substrates, and the potential connection of these features with magnetic ordering is less clear.

\clearpage

\textbf{S7. Characterization of hysteresis timescales}\\ 

\begin{figure*}[h]
\includegraphics[width=5 in]{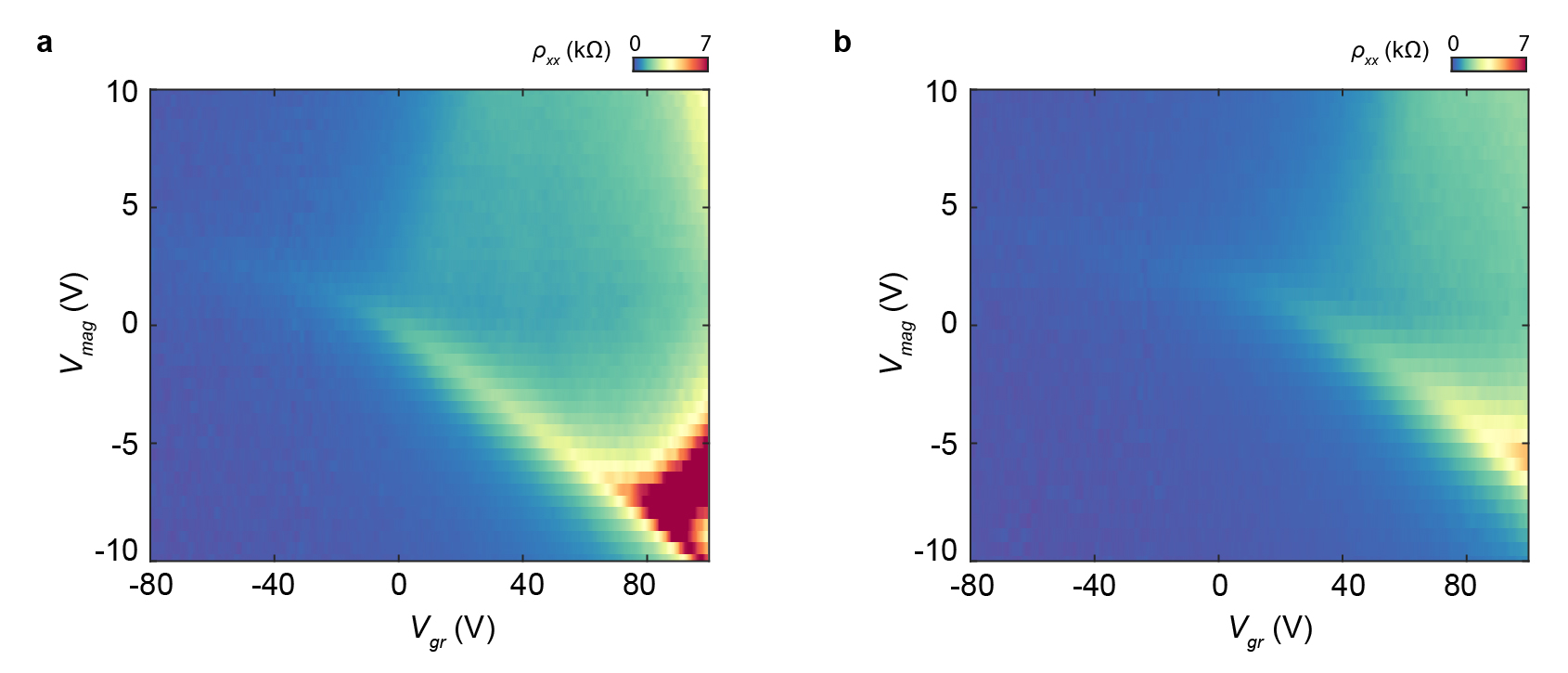} 
\caption{\textbf{Drifting of the charge doping in graphene over multiple days in Device A.}
\textbf{a-b}, Maps of $\rho_{xx}$ acquired by sweeping both gates at $B=0$~T. The measurements in \textbf{a} and \textbf{b} were performed three days apart within the same cooldown, under otherwise identical conditions.
}
\label{fig:long_drift}
\end{figure*}

\begin{figure*}[h]
\includegraphics[width=3.5 in]{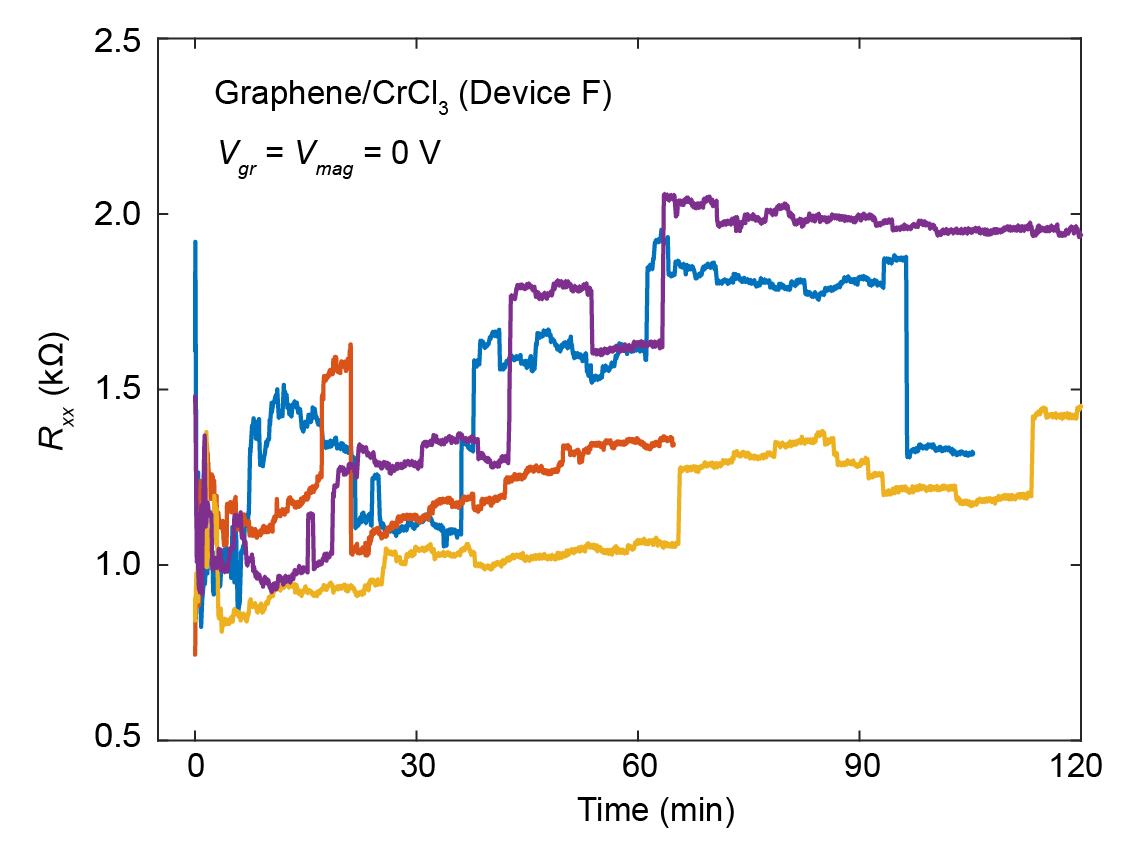} 
\caption{\textbf{Telegraph noise in the resistance of graphene on CrCl$_3$.}
Measurements of the graphene resistance in Device F acquired as a function of time with $V_{gr}=V_{mag}=0$ at $B=12$~T. The system is initialized by quickly sweeping from $V_{gr}=10$~V just prior to the measurement. The measurement is repeated four consecutive times, with each shown in a different color.
}
\label{fig:telegraph}
\end{figure*}

We routinely observe hysteresis over long time scales in our graphene/CrX$_3$ devices. Figures~\ref{fig:long_drift} and~\ref{fig:telegraph} illustrate this from two different perspectives. Figure~\ref{fig:long_drift} shows two identical $\rho_{xx}$ dual gate maps at $B=0$ acquired three days apart in Device A (but within the same cooldown). All of the features rigidly shift towards more positive values of $V_{gr}$. In particular, the Dirac point drifts almost entirely outside of the accessible gate range. The precise origin of this long-term drifting feature is not known, but must be related to slow charging effects of the CrI$_3$.

Figure~\ref{fig:telegraph} shows $\rho_{xx}$ in a graphene/CrCl$_3$ device (Device F) measured as a function of time at $B=12$~T. The measurement is initialized by quickly sweeping $V_{gr}$ from 10 V to 0 V with $V_{mag}=0$, and then waiting at $V_{gr}=0$~V while recording the device resistivity. Four consecutive measurements are performed in order to understand the reproducibility of the results, each shown in a different color. We see that the resistivity drifts upwards in all measurements over the span of at least two hours. We also observe large, abrupt jumps in the resistivity to both larger and smaller values at apparently random times. This telegraph noise indicates spontaneous mesoscopic rearrangement of the nearly-localized charges in the CrCl$_3$. Together, these measurements indicate the inability of the electrons to reach equilibrium in the CrX$_3$ substrates.

\clearpage

\textbf{S8. Transport in a graphene/WSe$_2$/CrI$_3$ heterostructure}\\

\begin{figure*}[h]
\includegraphics[width=6.9 in]{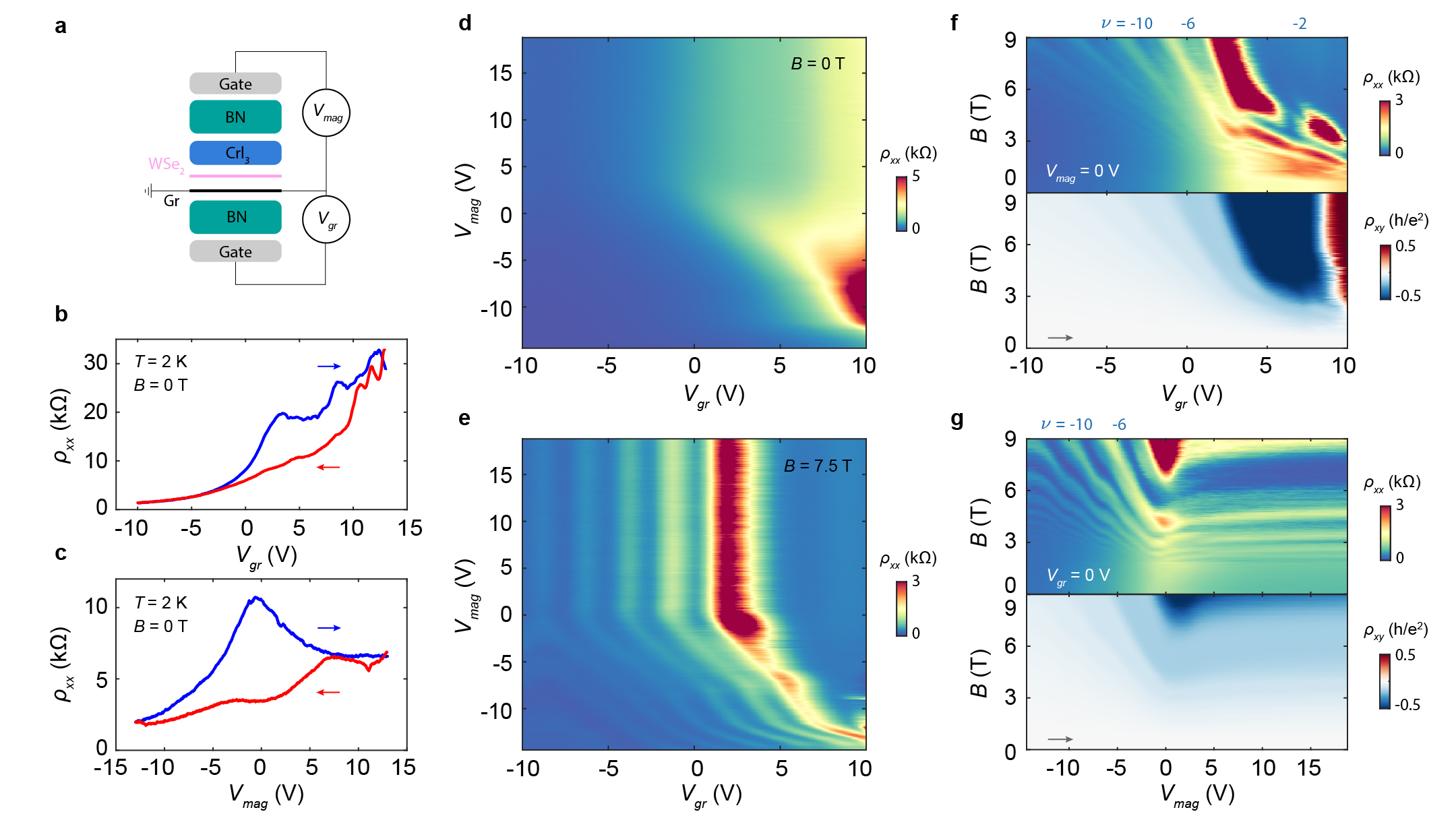} 
\caption{\textbf{Transport measurements in 3 layer CrI$_3$/monolayer WSe$_2$/graphene device (Device D).}
\textbf{a}, Cartoon schematic of the device structure. \textbf{b}, $\rho_{xx}$ as $V_{gr}$ is swept back and forth
with $V_{mag}=0$~V. \textbf{c}, $\rho_{xx}$ as $V_{mag}$ is swept back and forth with $V_{gr}=0$~V.
\textbf{d-e}, Maps of $\rho_{xx}$ acquired by sweeping both gates at $B=0$~T (\textbf{d}) and $B=7.5$~T (\textbf{e}), respectively. 
\textbf{f-g}, Landau fan diagrams acquired by sweeping $V_{gr}$ (\textbf{f}) and $V_{mag}$ (\textbf{g}), respectively, from negative to positive with the other gate held at ground.
}
\label{fig:CrI3_WSe2}
\end{figure*}

\begin{figure*}[t]
\includegraphics[width=3.5 in]{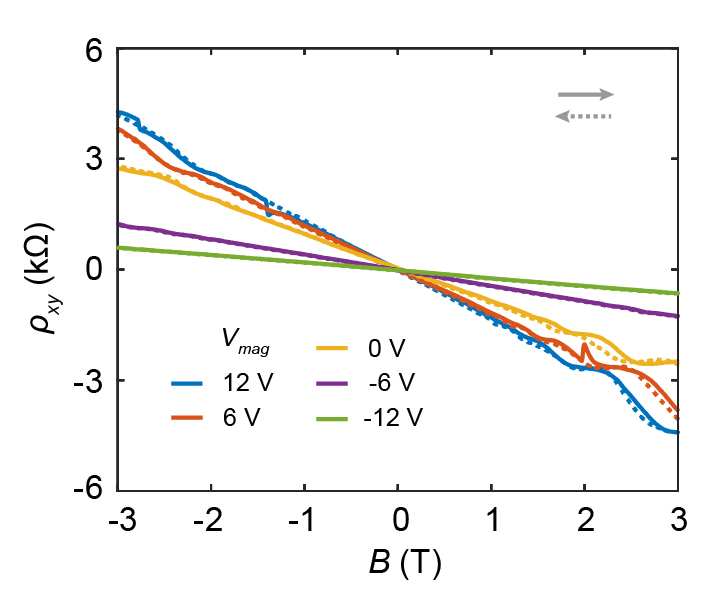} 
\caption{\textbf{Hall measurement in Device D.}
$\rho_{xy}$ measured as $B$ is swept back and forth with $V_{gr}=0$~V at different $V_{mag}$. Solid (dashed) lines denote the forward (backward) sweeping direction.
}
\label{fig:WSe2_Hall}
\end{figure*}

Interfacing graphene with WSe$_2$ (and similar transition metal dichalcogenides) is known enhance the spin-orbit coupling of the graphene~\cite{Avsar2014NatCom,Wang2015NatCom,Wang2016PRX,Yang2016SOC,Island2019}. We have additionally fabricated devices in which there is a monolayer of WSe$_2$ in between the graphene and the CrI$_3$. We find that even with the monolayer WSe$_2$ spacer layer, there is similar modulation doping of the graphene due to charge transfer with the CrI$_3$. Figure~\ref{fig:CrI3_WSe2} summarizes our primary observations in this device. Overall, we find that transport in this structure closely resembles the graphene/CrI$_3$ devices. We observe an ordinary Hall effect even very near the Dirac point (Fig.~\ref{fig:WSe2_Hall}), indicating the absence of the AHE in this system.  

\clearpage

\textbf{S9. Sensing the surface magnetic ordering of bulk CrI$_3$ with graphene}\\

\begin{figure*}[h]
\includegraphics[width=6.9 in]{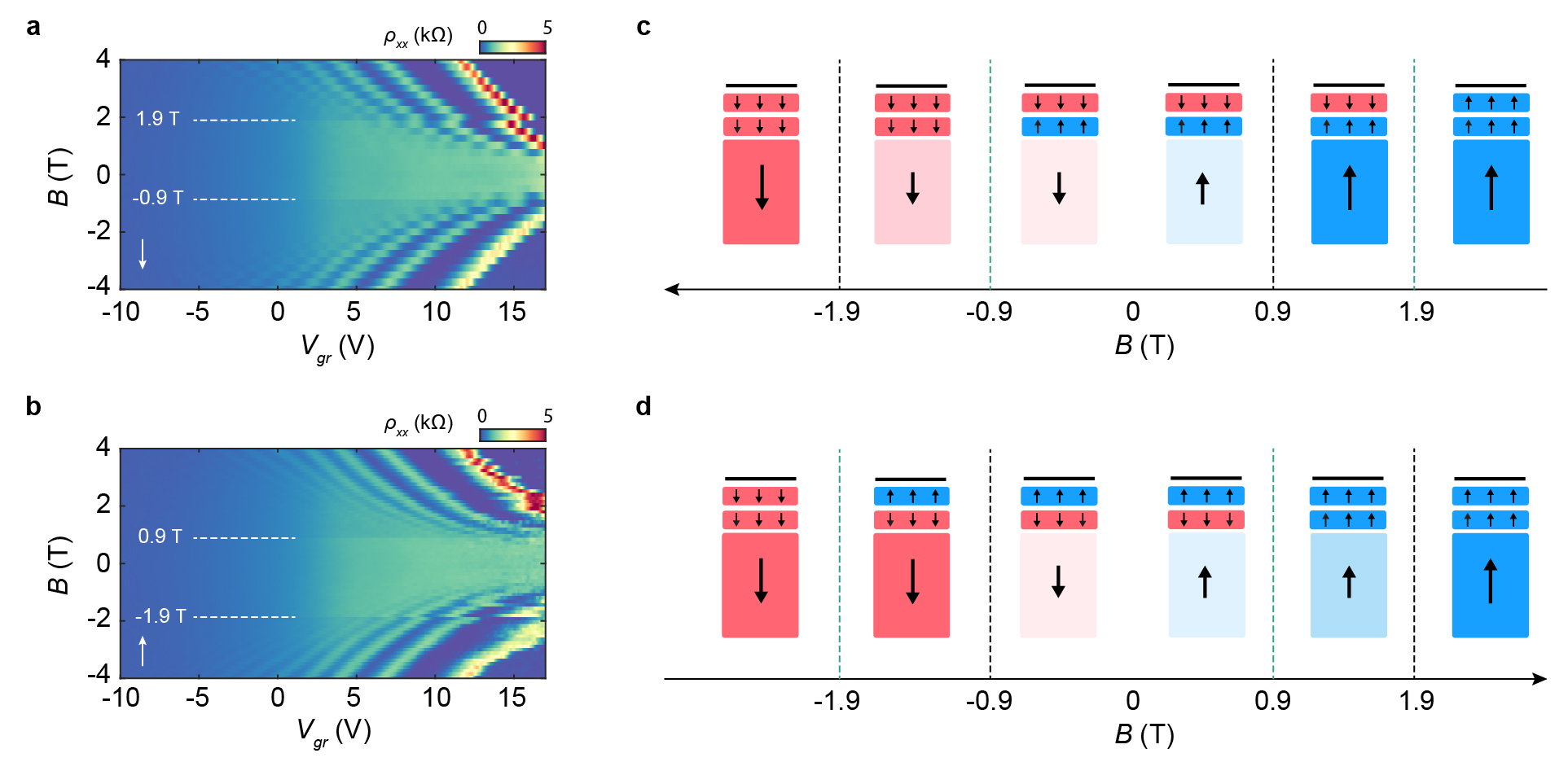} 
\caption{\textbf{Sensing the magnetic order of nearest few CrI$_3$ layers with graphene.}
\textbf{a-b}, Landau fan diagrams measured as the magnetic field is swept from positive to negative (\textbf{a}) and from negative to positive (\textbf{b}) in a device with graphene on a 10 layer CrI$_3$ (Device C). $V_{mag}=6$~V in both measurements. The white dashed lines denote the most robust abrupt jumps in the modulation doping of the graphene, determined as those which span the entire accessible range of $V_{gr}$. 
\textbf{c-d}, Schematics of the inferred magnetic ordering of the 10 layer CrI$_3$ as a function of $B$, for the case of sweeping $B$ from positive to negative (\textbf{c)} or visa versa (\textbf{d}). The graphene sheet is depicted as the solid black line atop a number of layers of CrI$_3$. Our measurements are evidently not sensitive to the spin configuration of the outer CrI$_3$ layers, which we aggregate into one large rectangle and color in light red/blue at intermediate magnetic fields to reflect a complicated (and unknown) interlayer magnetic ordering. Dashed green lines denote the value of $B$ corresponding to the experimentally observed jump in the graphene modulation doping, resulting from spin flips in either the nearest or next-nearest CrI$_3$ layer.
}
\label{fig:CrI3_collective}
\end{figure*}

Figure~\ref{fig:CrI3_collective}a (b) shows a Landau fan diagram of a graphene on 10 layer CrI$_3$ device (Device C) acquired by changing $V_{gr}$ at fixed $V_{mag}=6$~V, and with $B$ swept from positive to negative (negative to positive) values. We observe one abrupt jump in the modulation doping at all values of $V_{gr}$ for each sign of $B$, as indicated by the white dashed lines (other small jumps are also occasionally observed, but only occur over small ranges of $V_{gr}$ and likely arise due to randomly moving charges in the CrI$_3$). Unlike for the 3 layer CrI$_3$ device in which these jumps occur at the same absolute value of $B$, we find an asymmetry in the switching field in the 10 layer device. In particular, the first jump occurs when the field crosses $|B|=1.9$~T, and the second jump occurs when the opposite sign of the field exceeds $|B|=0.9$~T.

Magnetic switching at fields of 0.9~T and 1.9~T (in addition to at many smaller fields) is routinely observed in magnetic tunnel junction (MTJ) measurements of CrI$_3$ up to ten layers~\cite{Song2018SciCrI3MTJ,Song2019NanolettCrI3MTJ,Kim2018NanolettCrI3MTJ,Wang2018NatComCrI3MTJ,Kim2019PNASCrI3MTJ}. These structures, consisting of graphite/few-layer CrI$_3$/graphite, exhibit symmetric tunneling properties for both signs of the magnetic field. In contrast, our transport measurements are inherently asymmetric, since the graphene layer sits only on one side of the CrI$_3$. Seeing jumps in the modulation doping at different fields depending on the field sweeping direction indicates that the graphene is sensitive only to the spin configuration of the nearest few CrI$_3$ layers. The schematics in Figs.~\ref{fig:CrI3_collective}c-d illustrate one possible evolution of the interlayer spin arrangement as the magnetic field is swept. Starting from the interlayer ferromagnetic configuration at high field, our results are consistent with the spins in the CrI$_3$ layer neighboring the graphene flipping first at $|B|=1.9$~T. In contrast, the spins in the second layer do not flip until the field reaches a value of 0.9~T with the opposite sign. The spins in the remaining bulk layers likely flip in a correlated fashion, however, the graphene transport is apparently not sensitive to these spin flips.

\clearpage

\textbf{S10. Mobility of hole-type carriers in graphene on CrI$_3$}\\

\begin{figure*}[h]
\includegraphics[width=6.9 in]{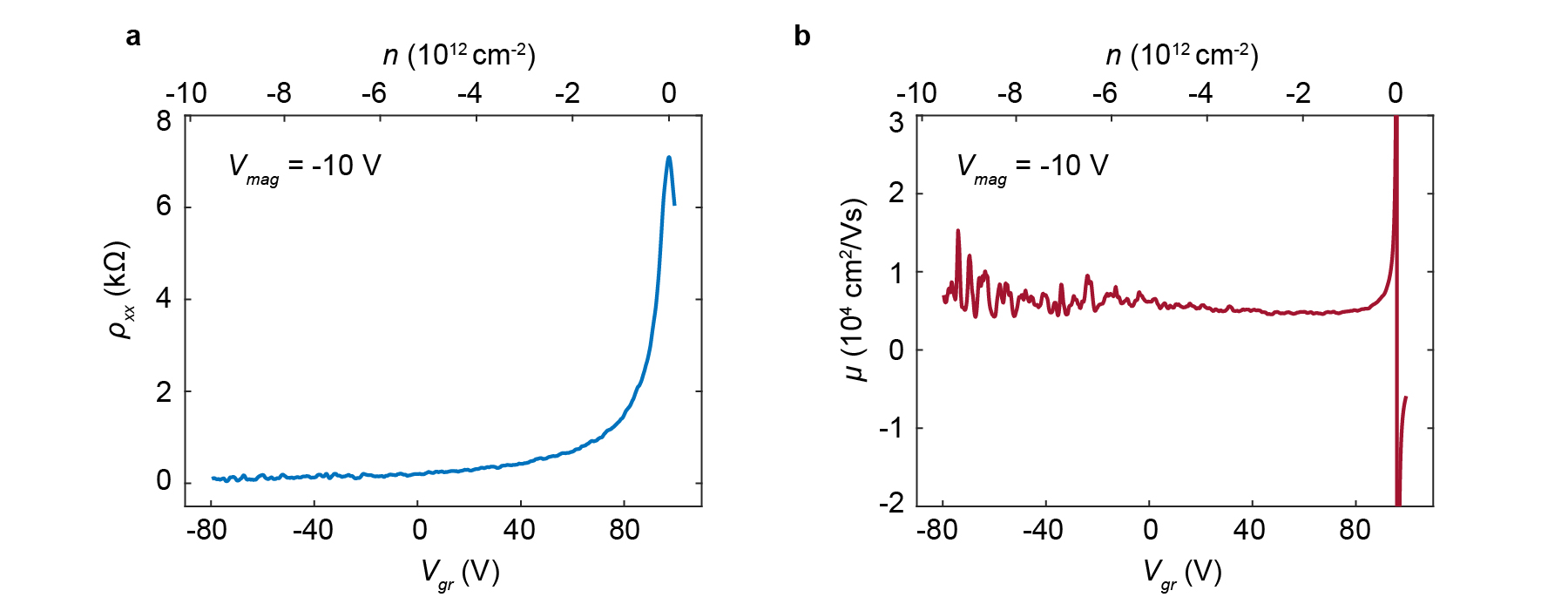} 
\caption{\textbf{Extraction of the graphene mobility in Device A.}
\textbf{a}, Graphene resistivity acquired as $V_{gr}$ is swept at $V_{mag}=-10$~V. The charge transfer is suppressed at this value of $V_{mag}$, therefore we are able to extract the charge carrier density, $n$, in the graphene from the known gate capacitance (top axis). \textbf{b}, Charge carrier mobility as a function of $V_{gr}$ (bottom axis) and $n$ (top axis), calculated as $\mu=1/e n \rho$.
}
\label{fig:mobility}
\end{figure*}

\end{document}